\documentclass[longauth]{aa}
\usepackage{newtxtext,newtxmath}
\usepackage{graphicx}
\usepackage{natbib}
\usepackage{scalerel}
\usepackage{comment}
\usepackage[T1]{fontenc}
\usepackage{amsmath}	% Advanced maths commands
\usepackage{multirow}
\usepackage{subcaption}
\usepackage{lastpage}

\usepackage{xcolor} % LW: I removed the 'table' option here as it appeared to be unneeded.

\usepackage[pdfencoding=auto,psdextra]{hyperref}
\hypersetup{
    colorlinks=true,
    linkcolor=blue,
    filecolor=magenta,      
    urlcolor=blue,
    citecolor=blue
}
\usepackage{orcidlink} % LW: I moved this to be after 'hyperref' to avoid conflict.

\makeatletter
\renewcommand*\aa@pageof{, page \thepage{} of \pageref*{LastPage}}
\newcommand{\ols}[1]{\mskip.5\thinmuskip\overline{\mskip-.5\thinmuskip {#1} \mskip-.5\thinmuskip}\mskip.5\thinmuskip} % overline short

\makeatother

\usepackage[utf8]{inputenc}

\usepackage{euclid}

\newcommand{\mathbfit}[1]{\textbf{\textit{#1}}}

\newcommand{\mathbfss}[1]{\textbf{\textsf{#1}}}

\newcommand{\nv}{\hat{\mathbfit{u}}}
\newcommand{\I}{{\rm i}}
\newcommand{\tp}{{\mkern-.5mu\mathsf{T}}}
\newcommand{\BinnedM}{\mathcal{M}}
\newcommand{\sqdeg}{\ensuremath{\,\mathrm{deg}^2}}
\newcommand{\orcid}[1]{\orcidlink{#1}} %% if already defined in aa.cls: comment, or use renewcommand
\newcommand{\estimator}[1]{\ensuremath{\widehat{#1}}}

\makeatletter
\def\@threej(#1,#2,#3,#4,#5,#6){%
\begin{pmatrix}#1&#3&#5\\#2&#4&#6\end{pmatrix}}
\newcommand{\threej}[3]{\@threej(#1,#2,#3)}
\makeatother

\iffalse
    \newcommand{\partfn}{f}
    \newcommand{\fullfn}{g}
    \newcommand{\partxi}{\xi^{f\!f'}}
    \newcommand{\fullxi}{\xi^{g\!g'}}
    \newcommand{\partC}[1]{C_{#1}^{ff'}} % Argument is the ell subscript
    \newcommand{\fullC}[1]{C_{#1}^{gg'}} % Argument is the ell subscript
\else
    \newcommand{\partfn}{\tilde{f}}
    \newcommand{\fullfn}{f}
    \newcommand{\partxi}{\tilde{\xi}^{f\!f'}}
    \newcommand{\fullxi}{\xi^{f\!f'}}
    \newcommand{\partC}[1]{\tilde{C}_{#1}^{ff'}} % Argument is the ell subscript
    \newcommand{\fullC}[1]{C_{#1}^{ff'}} % Argument is the ell subscript
    
\fi

\begin{document}
\title{\Euclid}
\subtitle{High signal-to-noise angular power spectrum reconstruction with limited sky coverage}
 %%%% Version Thursday 12th of November 2026 02:57:07 PM UT
%%%% Assumes the new A&A style file from Oct 2025 or later
%%%% Please do not edit the author list -- contact ECEB Bureau for changes	   
\author{Euclid Collaboration: J.~Ruiz-Zapatero\orcid{0000-0002-7951-4391}\thanks{\email{jaime.ruiz-zapatero@ucl.ac.uk}}\inst{\ref{aff1}}
\and N.~Tessore\orcid{0000-0002-9696-7931}\inst{\ref{aff2}}
\and L.~Whiteway\orcid{0000-0001-6375-0755}\inst{\ref{aff1}}
\and B.~Joachimi\orcid{0000-0001-7494-1303}\inst{\ref{aff1}}
\and N.~Jeffrey\orcid{0000-0003-2927-1800}\inst{\ref{aff1},\ref{aff3},\ref{aff4}}
\and M.~Gatti\inst{\ref{aff5}}
\and B.~Altieri\orcid{0000-0003-3936-0284}\inst{\ref{aff6}}
\and S.~Andreon\orcid{0000-0002-2041-8784}\inst{\ref{aff7}}
\and N.~Auricchio\orcid{0000-0003-4444-8651}\inst{\ref{aff8}}
\and C.~Baccigalupi\orcid{0000-0002-8211-1630}\inst{\ref{aff9},\ref{aff10},\ref{aff11},\ref{aff12}}
\and M.~Baldi\orcid{0000-0003-4145-1943}\inst{\ref{aff13},\ref{aff8},\ref{aff14}}
\and S.~Bardelli\orcid{0000-0002-8900-0298}\inst{\ref{aff8}}
\and P.~Battaglia\orcid{0000-0002-7337-5909}\inst{\ref{aff8}}
\and F.~Bernardeau\orcid{0009-0007-3015-2581}\inst{\ref{aff15},\ref{aff16}}
\and A.~Biviano\orcid{0000-0002-0857-0732}\inst{\ref{aff10},\ref{aff9}}
\and E.~Branchini\orcid{0000-0002-0808-6908}\inst{\ref{aff17},\ref{aff18},\ref{aff7}}
\and M.~Brescia\orcid{0000-0001-9506-5680}\inst{\ref{aff19},\ref{aff20}}
\and S.~Camera\orcid{0000-0003-3399-3574}\inst{\ref{aff21},\ref{aff22},\ref{aff23}}
\and G.~Ca\~nas-Herrera\orcid{0000-0003-2796-2149}\inst{\ref{aff24}}
\and V.~Capobianco\orcid{0000-0002-3309-7692}\inst{\ref{aff23}}
\and C.~Carbone\orcid{0000-0003-0125-3563}\inst{\ref{aff25}}
\and J.~Carretero\orcid{0000-0002-3130-0204}\inst{\ref{aff26},\ref{aff27}}
\and F.~J.~Castander\orcid{0000-0001-7316-4573}\inst{\ref{aff28},\ref{aff29}}
\and M.~Castellano\orcid{0000-0001-9875-8263}\inst{\ref{aff30}}
\and G.~Castignani\orcid{0000-0001-6831-0687}\inst{\ref{aff8}}
\and S.~Cavuoti\orcid{0000-0002-3787-4196}\inst{\ref{aff20},\ref{aff31}}
\and A.~Cimatti\inst{\ref{aff32}}
\and C.~Colodro-Conde\inst{\ref{aff33}}
\and G.~Congedo\orcid{0000-0003-2508-0046}\inst{\ref{aff34}}
\and L.~Conversi\orcid{0000-0002-6710-8476}\inst{\ref{aff35},\ref{aff6}}
\and Y.~Copin\orcid{0000-0002-5317-7518}\inst{\ref{aff36}}
\and F.~Courbin\orcid{0000-0003-0758-6510}\inst{\ref{aff37},\ref{aff38},\ref{aff39}}
\and H.~M.~Courtois\orcid{0000-0003-0509-1776}\inst{\ref{aff40}}
\and M.~Crocce\orcid{0000-0002-9745-6228}\inst{\ref{aff28},\ref{aff29}}
\and A.~Da~Silva\orcid{0000-0002-6385-1609}\inst{\ref{aff41},\ref{aff42}}
\and H.~Degaudenzi\orcid{0000-0002-5887-6799}\inst{\ref{aff43}}
\and G.~De~Lucia\orcid{0000-0002-6220-9104}\inst{\ref{aff10}}
\and H.~Dole\orcid{0000-0002-9767-3839}\inst{\ref{aff44}}
\and F.~Dubath\orcid{0000-0002-6533-2810}\inst{\ref{aff43}}
\and C.~A.~J.~Duncan\orcid{0009-0003-3573-0791}\inst{\ref{aff34}}
\and X.~Dupac\inst{\ref{aff6}}
\and S.~Escoffier\orcid{0000-0002-2847-7498}\inst{\ref{aff45}}
\and M.~Farina\orcid{0000-0002-3089-7846}\inst{\ref{aff46}}
\and R.~Farinelli\inst{\ref{aff8}}
\and S.~Farrens\orcid{0000-0002-9594-9387}\inst{\ref{aff47}}
\and F.~Faustini\orcid{0000-0001-6274-5145}\inst{\ref{aff30}}
\and S.~Ferriol\inst{\ref{aff36}}
\and F.~Finelli\orcid{0000-0002-6694-3269}\inst{\ref{aff8},\ref{aff48}}
\and P.~Fosalba\orcid{0000-0002-1510-5214}\inst{\ref{aff29},\ref{aff28}}
\and M.~Frailis\orcid{0000-0002-7400-2135}\inst{\ref{aff10}}
\and E.~Franceschi\orcid{0000-0002-0585-6591}\inst{\ref{aff8}}
\and M.~Fumana\orcid{0000-0001-6787-5950}\inst{\ref{aff25}}
\and L.~Gabarra\orcid{0000-0002-8486-8856}\inst{\ref{aff49}}
\and S.~Galeotta\orcid{0000-0002-3748-5115}\inst{\ref{aff10}}
\and K.~George\orcid{0000-0002-1734-8455}\inst{\ref{aff50}}
\and B.~Gillis\orcid{0000-0002-4478-1270}\inst{\ref{aff34}}
\and C.~Giocoli\orcid{0000-0002-9590-7961}\inst{\ref{aff8},\ref{aff14}}
\and J.~Gracia-Carpio\orcid{0000-0003-4689-3134}\inst{\ref{aff51}}
\and A.~Grazian\orcid{0000-0002-5688-0663}\inst{\ref{aff52}}
\and F.~Grupp\inst{\ref{aff51},\ref{aff53}}
\and W.~G.~Hartley\orcid{0000-0002-4563-5101}\inst{\ref{aff43}}
\and S.~V.~H.~Haugan\orcid{0000-0001-9648-7260}\inst{\ref{aff54}}
\and H.~Hoekstra\orcid{0000-0002-0641-3231}\inst{\ref{aff24}}
\and W.~Holmes\orcid{0009-0007-8554-4646}\inst{\ref{aff55}}
\and F.~Hormuth\inst{\ref{aff56}}
\and A.~Hornstrup\orcid{0000-0002-3363-0936}\inst{\ref{aff57},\ref{aff58}}
\and K.~Jahnke\orcid{0000-0003-3804-2137}\inst{\ref{aff59}}
\and M.~Jhabvala\inst{\ref{aff60}}
\and S.~Kermiche\orcid{0000-0002-0302-5735}\inst{\ref{aff45}}
\and A.~Kiessling\orcid{0000-0002-2590-1273}\inst{\ref{aff55}}
\and B.~Kubik\orcid{0009-0006-5823-4880}\inst{\ref{aff36}}
\and M.~Kunz\orcid{0000-0002-3052-7394}\inst{\ref{aff61}}
\and H.~Kurki-Suonio\orcid{0000-0002-4618-3063}\inst{\ref{aff62},\ref{aff63}}
\and A.~M.~C.~Le~Brun\orcid{0000-0002-0936-4594}\inst{\ref{aff64}}
\and S.~Ligori\orcid{0000-0003-4172-4606}\inst{\ref{aff23}}
\and P.~B.~Lilje\orcid{0000-0003-4324-7794}\inst{\ref{aff54}}
\and V.~Lindholm\orcid{0000-0003-2317-5471}\inst{\ref{aff62},\ref{aff63}}
\and I.~Lloro\orcid{0000-0001-5966-1434}\inst{\ref{aff65}}
\and M.~Magliocchetti\orcid{0000-0001-9158-4838}\inst{\ref{aff46}}
\and G.~Mainetti\orcid{0000-0003-2384-2377}\inst{\ref{aff66}}
\and O.~Mansutti\orcid{0000-0001-5758-4658}\inst{\ref{aff10}}
\and O.~Marggraf\orcid{0000-0001-7242-3852}\inst{\ref{aff67}}
\and M.~Martinelli\orcid{0000-0002-6943-7732}\inst{\ref{aff30},\ref{aff68}}
\and N.~Martinet\orcid{0000-0003-2786-7790}\inst{\ref{aff69}}
\and F.~Marulli\orcid{0000-0002-8850-0303}\inst{\ref{aff70},\ref{aff8},\ref{aff14}}
\and R.~J.~Massey\orcid{0000-0002-6085-3780}\inst{\ref{aff71}}
\and N.~Mauri\orcid{0000-0001-8196-1548}\inst{\ref{aff32},\ref{aff14}}
\and E.~Medinaceli\orcid{0000-0002-4040-7783}\inst{\ref{aff8}}
\and S.~Mei\orcid{0000-0002-2849-559X}\inst{\ref{aff72},\ref{aff73}}
\and M.~Meneghetti\orcid{0000-0003-1225-7084}\inst{\ref{aff8},\ref{aff14}}
\and E.~Merlin\orcid{0000-0001-6870-8900}\inst{\ref{aff52}}
\and G.~Meylan\orcid{0000-0001-6503-0209}\inst{\ref{aff74}}
\and P.~Monaco\orcid{0000-0003-2083-7564}\inst{\ref{aff75},\ref{aff10},\ref{aff11},\ref{aff9}}
\and A.~Mora\orcid{0000-0002-1922-8529}\inst{\ref{aff76}}
\and M.~Moresco\orcid{0000-0002-7616-7136}\inst{\ref{aff70},\ref{aff8}}
\and C.~Moretti\orcid{0000-0003-3314-8936}\inst{\ref{aff10},\ref{aff9},\ref{aff11}}
\and L.~Moscardini\orcid{0000-0002-3473-6716}\inst{\ref{aff70},\ref{aff8},\ref{aff14}}
\and C.~Neissner\orcid{0000-0001-8524-4968}\inst{\ref{aff77},\ref{aff27}}
\and S.-M.~Niemi\orcid{0009-0005-0247-0086}\inst{\ref{aff78}}
\and C.~Padilla\orcid{0000-0001-7951-0166}\inst{\ref{aff77}}
\and S.~Paltani\orcid{0000-0002-8108-9179}\inst{\ref{aff43}}
\and F.~Pasian\orcid{0000-0002-4869-3227}\inst{\ref{aff10}}
\and V.~Pettorino\orcid{0000-0002-4203-9320}\inst{\ref{aff78}}
\and A.~Pezzotta\orcid{0000-0003-0726-2268}\inst{\ref{aff7}}
\and S.~Pires\orcid{0000-0002-0249-2104}\inst{\ref{aff47}}
\and G.~Polenta\orcid{0000-0003-4067-9196}\inst{\ref{aff79}}
\and M.~Poncet\inst{\ref{aff80}}
\and L.~A.~Popa\inst{\ref{aff81}}
\and F.~Raison\orcid{0000-0002-7819-6918}\inst{\ref{aff51}}
\and A.~Renzi\orcid{0000-0001-9856-1970}\inst{\ref{aff82},\ref{aff83},\ref{aff8}}
\and J.~Rhodes\orcid{0000-0002-4485-8549}\inst{\ref{aff55}}
\and G.~Riccio\inst{\ref{aff20}}
\and I.~Risso\orcid{0000-0003-2525-7761}\inst{\ref{aff17},\ref{aff18},\ref{aff7}}
\and F.~Rizzo\orcid{0000-0002-9407-585X}\inst{\ref{aff10}}
\and E.~Romelli\orcid{0000-0003-3069-9222}\inst{\ref{aff10}}
\and M.~Roncarelli\orcid{0000-0001-9587-7822}\inst{\ref{aff8}}
\and R.~Saglia\orcid{0000-0003-0378-7032}\inst{\ref{aff53},\ref{aff51}}
\and Z.~Sakr\orcid{0000-0002-4823-3757}\inst{\ref{aff84},\ref{aff85},\ref{aff86}}
\and D.~Sapone\orcid{0000-0001-7089-4503}\inst{\ref{aff87}}
\and B.~Sartoris\orcid{0000-0003-1337-5269}\inst{\ref{aff51},\ref{aff10}}
\and P.~Schneider\orcid{0000-0001-8561-2679}\inst{\ref{aff67}}
\and A.~Secroun\orcid{0000-0003-0505-3710}\inst{\ref{aff45}}
\and E.~Sihvola\orcid{0000-0003-1804-7715}\inst{\ref{aff88}}
\and P.~Simon\inst{\ref{aff67}}
\and C.~Sirignano\orcid{0000-0002-0995-7146}\inst{\ref{aff82},\ref{aff83}}
\and G.~Sirri\orcid{0000-0003-2626-2853}\inst{\ref{aff14}}
\and A.~Spurio~Mancini\orcid{0000-0001-5698-0990}\inst{\ref{aff89}}
\and L.~Stanco\orcid{0000-0002-9706-5104}\inst{\ref{aff83}}
\and P.~Tallada-Cresp\'{i}\orcid{0000-0002-1336-8328}\inst{\ref{aff26},\ref{aff27}}
\and A.~N.~Taylor\inst{\ref{aff34}}
\and I.~Tereno\orcid{0000-0002-4537-6218}\inst{\ref{aff41},\ref{aff90}}
\and S.~Toft\orcid{0000-0003-3631-7176}\inst{\ref{aff91},\ref{aff92}}
\and R.~Toledo-Moreo\orcid{0000-0002-2997-4859}\inst{\ref{aff93},\ref{aff94}}
\and F.~Torradeflot\orcid{0000-0003-1160-1517}\inst{\ref{aff27},\ref{aff26}}
\and I.~Tutusaus\orcid{0000-0002-3199-0399}\inst{\ref{aff28},\ref{aff29},\ref{aff85}}
\and J.~Valiviita\orcid{0000-0001-6225-3693}\inst{\ref{aff62},\ref{aff63}}
\and T.~Vassallo\orcid{0000-0001-6512-6358}\inst{\ref{aff10},\ref{aff50}}
\and A.~Veropalumbo\orcid{0000-0003-2387-1194}\inst{\ref{aff7},\ref{aff18},\ref{aff17}}
\and Y.~Wang\orcid{0000-0002-4749-2984}\inst{\ref{aff95}}
\and J.~Weller\orcid{0000-0002-8282-2010}\inst{\ref{aff53},\ref{aff51}}
\and A.~Zacchei\orcid{0000-0003-0396-1192}\inst{\ref{aff10},\ref{aff9}}
\and G.~Zamorani\orcid{0000-0002-2318-301X}\inst{\ref{aff8}}
\and F.~M.~Zerbi\orcid{0000-0002-9996-973X}\inst{\ref{aff7}}
\and E.~Zucca\orcid{0000-0002-5845-8132}\inst{\ref{aff8}}
\and M.~Ballardini\orcid{0000-0003-4481-3559}\inst{\ref{aff96},\ref{aff97},\ref{aff8}}
\and P.~Bergamini\orcid{0000-0003-1383-9414}\inst{\ref{aff8}}
\and S.~Borgani\orcid{0000-0001-6151-6439}\inst{\ref{aff75},\ref{aff9},\ref{aff10},\ref{aff11},\ref{aff98}}
\and E.~Bozzo\orcid{0000-0002-8201-1525}\inst{\ref{aff43}}
\and C.~Burigana\orcid{0000-0002-3005-5796}\inst{\ref{aff99},\ref{aff48}}
\and R.~Cabanac\orcid{0000-0001-6679-2600}\inst{\ref{aff85}}
\and M.~Calabrese\orcid{0000-0002-2637-2422}\inst{\ref{aff100},\ref{aff25}}
\and A.~Cappi\inst{\ref{aff101},\ref{aff8}}
\and F.~Caro\orcid{0009-0003-1053-0507}\inst{\ref{aff30}}
\and T.~Castro\orcid{0000-0002-6292-3228}\inst{\ref{aff102},\ref{aff10},\ref{aff9}}
\and D.~Di~Ferdinando\inst{\ref{aff14}}
\and J.~A.~Escartin~Vigo\inst{\ref{aff51}}
\and G.~Fabbian\orcid{0000-0002-3255-4695}\inst{\ref{aff44}}
\and J.~Garc\'ia-Bellido\orcid{0000-0002-9370-8360}\inst{\ref{aff84}}
\and V.~Gautard\inst{\ref{aff103}}
\and C.~M.~Gutierrez\orcid{0000-0001-7854-783X}\inst{\ref{aff33},\ref{aff104}}
\and A.~Hall\orcid{0000-0002-3139-8651}\inst{\ref{aff34}}
\and A.~Loureiro\orcid{0000-0002-4371-0876}\inst{\ref{aff105},\ref{aff106}}
\and J.~Macias-Perez\orcid{0000-0002-5385-2763}\inst{\ref{aff107}}
\and R.~Maoli\orcid{0000-0002-6065-3025}\inst{\ref{aff108},\ref{aff30}}
\and R.~B.~Metcalf\orcid{0000-0003-3167-2574}\inst{\ref{aff70},\ref{aff8}}
\and M.~P\"ontinen\orcid{0000-0001-5442-2530}\inst{\ref{aff62}}
\and E.~Sarpa\orcid{0000-0002-1256-655X}\inst{\ref{aff10}}
\and V.~Scottez\orcid{0009-0008-3864-940X}\inst{\ref{aff109},\ref{aff110}}
\and M.~Sereno\orcid{0000-0003-0302-0325}\inst{\ref{aff8},\ref{aff14}}
\and K.~Tanidis\orcid{0000-0001-9843-5130}\inst{\ref{aff111}}
\and M.~Tenti\orcid{0000-0002-4254-5901}\inst{\ref{aff14}}
\and M.~Tucci\inst{\ref{aff43}}
\and M.~Viel\orcid{0000-0002-2642-5707}\inst{\ref{aff9},\ref{aff10},\ref{aff12},\ref{aff11},\ref{aff98}}
\and M.~Wiesmann\orcid{0009-0000-8199-5860}\inst{\ref{aff54}}
\and J.~A.~Acevedo~Barroso\orcid{0000-0002-9654-1711}\inst{\ref{aff55}}
\and Y.~Akrami\orcid{0000-0002-2407-7956}\inst{\ref{aff84},\ref{aff112}}
\and I.~T.~Andika\orcid{0000-0001-6102-9526}\inst{\ref{aff53}}
\and G.~Angora\orcid{0000-0002-0316-6562}\inst{\ref{aff20},\ref{aff96}}
\and S.~Anselmi\orcid{0000-0002-3579-9583}\inst{\ref{aff83},\ref{aff82},\ref{aff113}}
\and M.~Archidiacono\orcid{0000-0003-4952-9012}\inst{\ref{aff114},\ref{aff115}}
\and G.~Aric\`o\orcid{0000-0002-2802-2928}\inst{\ref{aff14}}
\and F.~Atrio-Barandela\orcid{0000-0002-2130-2513}\inst{\ref{aff116}}
\and M.~Baes\orcid{0000-0002-3930-2757}\inst{\ref{aff117}}
\and L.~Bazzanini\orcid{0000-0003-0727-0137}\inst{\ref{aff96},\ref{aff8}}
\and D.~Bertacca\orcid{0000-0002-2490-7139}\inst{\ref{aff82},\ref{aff52},\ref{aff83}}
\and M.~Bethermin\orcid{0000-0002-3915-2015}\inst{\ref{aff118}}
\and F.~Beutler\orcid{0000-0003-0467-5438}\inst{\ref{aff34}}
\and A.~Blanchard\orcid{0000-0001-8555-9003}\inst{\ref{aff85}}
\and L.~Blot\orcid{0000-0002-9622-7167}\inst{\ref{aff119},\ref{aff64}}
\and M.~Bonici\orcid{0000-0002-8430-126X}\inst{\ref{aff120},\ref{aff25}}
\and F.~Bouch\`e\orcid{0000-0002-4663-1786}\inst{\ref{aff121},\ref{aff31}}
\and M.~L.~Brown\orcid{0000-0002-0370-8077}\inst{\ref{aff122}}
\and S.~Bruton\orcid{0000-0002-6503-5218}\inst{\ref{aff123}}
\and A.~Calabro\orcid{0000-0003-2536-1614}\inst{\ref{aff30}}
\and B.~Camacho~Quevedo\orcid{0000-0002-8789-4232}\inst{\ref{aff9},\ref{aff12},\ref{aff10}}
\and P.~Carrilho\orcid{0000-0003-1339-0194}\inst{\ref{aff124}}
\and C.~S.~Carvalho\inst{\ref{aff90}}
\and F.~Cogato\orcid{0000-0003-4632-6113}\inst{\ref{aff8}}
\and T.~E.~Collett\orcid{0000-0001-5564-3140}\inst{\ref{aff125}}
\and A.~R.~Cooray\orcid{0000-0002-3892-0190}\inst{\ref{aff126}}
\and P.~Corcho-Caballero\orcid{0000-0001-6327-7080}\inst{\ref{aff127}}
\and M.~Correnti\orcid{0000-0001-6464-3257}\inst{\ref{aff30},\ref{aff79}}
\and B.~Csizi\orcid{0000-0003-3227-6581}\inst{\ref{aff128}}
\and T.~de~Boer\orcid{0000-0001-5486-2747}\inst{\ref{aff129}}
\and G.~Desprez\orcid{0000-0001-8325-1742}\inst{\ref{aff127}}
\and A.~D\'iaz-S\'anchez\orcid{0000-0003-0748-4768}\inst{\ref{aff130}}
\and S.~Di~Domizio\orcid{0000-0003-2863-5895}\inst{\ref{aff17},\ref{aff18}}
\and J.~M.~Diego\orcid{0000-0001-9065-3926}\inst{\ref{aff131}}
\and V.~Duret\orcid{0009-0009-0383-4960}\inst{\ref{aff45}}
\and M.~Y.~Elkhashab\orcid{0000-0001-9306-2603}\inst{\ref{aff10},\ref{aff11},\ref{aff75},\ref{aff9}}
\and Y.~Fang\orcid{0000-0002-0334-6950}\inst{\ref{aff53}}
\and A.~Farina\orcid{0009-0000-3420-929X}\inst{\ref{aff7},\ref{aff18}}
\and A.~Finoguenov\orcid{0000-0002-4606-5403}\inst{\ref{aff62}}
\and A.~Franco\orcid{0000-0002-4761-366X}\inst{\ref{aff132},\ref{aff133},\ref{aff134}}
\and A.~Fumagalli\orcid{0009-0004-0300-2535}\inst{\ref{aff10}}
\and Y.~Fu\orcid{0000-0002-0759-0504}\inst{\ref{aff24},\ref{aff127}}
\and K.~Ganga\orcid{0000-0001-8159-8208}\inst{\ref{aff72}}
\and E.~Gaztanaga\orcid{0000-0001-9632-0815}\inst{\ref{aff28},\ref{aff29},\ref{aff125}}
\and Z.~Ghaffari\orcid{0000-0002-6467-8078}\inst{\ref{aff10},\ref{aff9}}
\and F.~Giacomini\orcid{0000-0002-3129-2814}\inst{\ref{aff14}}
\and F.~Gianotti\orcid{0000-0003-4666-119X}\inst{\ref{aff8}}
\and G.~Gozaliasl\orcid{0000-0002-0236-919X}\inst{\ref{aff135},\ref{aff62}}
\and A.~Gruppuso\orcid{0000-0001-9272-5292}\inst{\ref{aff8},\ref{aff14}}
\and M.~Guidi\orcid{0000-0001-9408-1101}\inst{\ref{aff13},\ref{aff8}}
\and C.~Hern\'andez-Monteagudo\orcid{0000-0001-5471-9166}\inst{\ref{aff104},\ref{aff33}}
\and H.~Hildebrandt\orcid{0000-0002-9814-3338}\inst{\ref{aff136}}
\and J.~Hjorth\orcid{0000-0002-4571-2306}\inst{\ref{aff137}}
\and M.~Jauzac\orcid{0000-0003-1974-8732}\inst{\ref{aff138},\ref{aff71},\ref{aff139},\ref{aff140},\ref{aff85}}
\and S.~Joudaki\orcid{0000-0001-8820-673X}\inst{\ref{aff26},\ref{aff125}}
\and J.~J.~E.~Kajava\orcid{0000-0002-3010-8333}\inst{\ref{aff141},\ref{aff142},\ref{aff143}}
\and Y.~Kang\orcid{0009-0000-8588-7250}\inst{\ref{aff128}}
\and V.~Kansal\orcid{0000-0002-4008-6078}\inst{\ref{aff144},\ref{aff145}}
\and D.~Karagiannis\orcid{0000-0002-4927-0816}\inst{\ref{aff96},\ref{aff146}}
\and J.~Kim\orcid{0000-0003-2776-2761}\inst{\ref{aff49}}
\and C.~C.~Kirkpatrick\inst{\ref{aff88}}
\and A.~Kov\'acs\orcid{0000-0002-5825-579X}\inst{\ref{aff147},\ref{aff148}}
\and I.~Kova{\v{c}}i{\'{c}}\orcid{0000-0001-6751-3263}\inst{\ref{aff117}}
\and K.~Koyama\orcid{0000-0001-6727-6915}\inst{\ref{aff125}}
\and S.~Kruk\orcid{0000-0001-8010-8879}\inst{\ref{aff6}}
\and M.~C.~Lam\orcid{0000-0002-9347-2298}\inst{\ref{aff34}}
\and M.~Lattanzi\orcid{0000-0003-1059-2532}\inst{\ref{aff97}}
\and F.~Leclercq\orcid{0000-0002-9339-1404}\inst{\ref{aff16}}
\and S.~Lee\orcid{0000-0002-8289-740X}\inst{\ref{aff149},\ref{aff55}}
\and L.~Legrand\orcid{0000-0003-0610-5252}\inst{\ref{aff150},\ref{aff44}}
\and M.~Lembo\orcid{0000-0002-5271-5070}\inst{\ref{aff16}}
\and G.~Leroy\orcid{0009-0004-2523-4425}\inst{\ref{aff138},\ref{aff71}}
\and J.~Lesgourgues\orcid{0000-0001-7627-353X}\inst{\ref{aff151}}
\and L.~Linke\orcid{0000-0002-2622-8113}\inst{\ref{aff128}}
\and S.~J.~Liu\orcid{0000-0001-7680-2139}\inst{\ref{aff46}}
\and G.~Maggio\orcid{0000-0003-4020-4836}\inst{\ref{aff10}}
\and F.~Mannucci\orcid{0000-0002-4803-2381}\inst{\ref{aff152}}
\and F.~R.~Marleau\orcid{0000-0002-1442-2947}\inst{\ref{aff128}}
\and C.~J.~A.~P.~Martins\orcid{0000-0002-4886-9261}\inst{\ref{aff153},\ref{aff154}}
\and M.~Migliaccio\inst{\ref{aff155},\ref{aff156}}
\and M.~Miluzio\inst{\ref{aff6},\ref{aff157}}
\and G.~Morgante\inst{\ref{aff8}}
\and S.~Nadathur\orcid{0000-0001-9070-3102}\inst{\ref{aff125}}
\and K.~Naidoo\orcid{0000-0002-9182-1802}\inst{\ref{aff125},\ref{aff59}}
\and P.~Natoli\orcid{0000-0003-0126-9100}\inst{\ref{aff96},\ref{aff97}}
\and A.~Navarro-Alsina\orcid{0000-0002-3173-2592}\inst{\ref{aff67}}
\and S.~Nesseris\orcid{0000-0002-0567-0324}\inst{\ref{aff84}}
\and L.~Nicastro\orcid{0000-0001-8534-6788}\inst{\ref{aff8}}
\and F.~Oppizzi\orcid{0000-0003-3904-8370}\inst{\ref{aff18},\ref{aff83},\ref{aff82}}
\and F.~Pace\orcid{0000-0001-8039-0480}\inst{\ref{aff21},\ref{aff22},\ref{aff23},\ref{aff42}}
\and D.~Paoletti\orcid{0000-0003-4761-6147}\inst{\ref{aff8},\ref{aff48}}
\and F.~Passalacqua\orcid{0000-0002-8606-4093}\inst{\ref{aff83}}
\and K.~Paterson\orcid{0000-0001-8340-3486}\inst{\ref{aff59}}
\and C.~Pattison\orcid{0000-0003-3272-2617}\inst{\ref{aff125}}
\and A.~Pisani\orcid{0000-0002-6146-4437}\inst{\ref{aff45}}
\and L.~Pizzuti\orcid{0000-0001-5654-7580}\inst{\ref{aff158}}
\and D.~Potter\orcid{0000-0002-0757-5195}\inst{\ref{aff159}}
\and G.~W.~Pratt\inst{\ref{aff47}}
\and S.~Quai\orcid{0000-0002-0449-8163}\inst{\ref{aff70},\ref{aff8}}
\and M.~Radovich\orcid{0000-0002-3585-866X}\inst{\ref{aff52}}
\and W.~Roster\orcid{0000-0002-9149-6528}\inst{\ref{aff51}}
\and M.~Sahl\'en\orcid{0000-0003-0973-4804}\inst{\ref{aff160}}
\and D.~B.~Sanders\orcid{0000-0002-1233-9998}\inst{\ref{aff129}}
\and A.~Schneider\orcid{0000-0001-7055-8104}\inst{\ref{aff159}}
\and D.~Sciotti\orcid{0009-0008-4519-2620}\inst{\ref{aff30},\ref{aff68}}
\and E.~Sellentin\orcid{0009-0002-2655-3458}\inst{\ref{aff161},\ref{aff24}}
\and L.~C.~Smith\orcid{0000-0002-3259-2771}\inst{\ref{aff162}}
\and I.~Szapudi\orcid{0000-0003-2274-0301}\inst{\ref{aff129}}
\and F.~Tarsitano\orcid{0000-0002-5919-0238}\inst{\ref{aff163},\ref{aff164}}
\and G.~Testera\orcid{0000-0003-2970-766X}\inst{\ref{aff18}}
\and R.~Teyssier\orcid{0000-0001-7689-0933}\inst{\ref{aff165}}
\and S.~Tosi\orcid{0000-0002-7275-9193}\inst{\ref{aff17},\ref{aff7},\ref{aff18}}
\and A.~Troja\orcid{0000-0003-0239-4595}\inst{\ref{aff10}}
\and C.~Uhlemann\orcid{0000-0001-7831-1579}\inst{\ref{aff166},\ref{aff167}}
\and C.~Valieri\inst{\ref{aff14}}
\and D.~Vergani\orcid{0000-0003-0898-2216}\inst{\ref{aff8}}
\and P.~Vielzeuf\orcid{0000-0003-2035-9339}\inst{\ref{aff45}}
\and S.~Vinciguerra\orcid{0009-0005-4018-3184}\inst{\ref{aff69}}
\and M.~von~Wietersheim-Kramsta\orcid{0000-0003-4986-5091}\inst{\ref{aff71},\ref{aff138}}
\and N.~A.~Walton\orcid{0000-0003-3983-8778}\inst{\ref{aff162}}
\and L.~Wang\orcid{0000-0002-6736-9158}\inst{\ref{aff168},\ref{aff127}}
\and A.~H.~Wright\orcid{0000-0001-7363-7932}\inst{\ref{aff136}}}
										   
%%%% please do not edit the affiliation list -- contact ECEB Bureau for changes
\institute{Department of Physics and Astronomy, University College London, Gower Street, London WC1E 6BT, UK\label{aff1}
\and
Mullard Space Science Laboratory, University College London, Holmbury St Mary, Dorking, Surrey RH5 6NT, UK\label{aff2}
\and
Department of Physics, King's College London, Strand, London, WC2R 2LS, UK\label{aff3}
\and
King's Institute for Artificial Intelligence, King's College London, Strand, London, WC2R 2LS, UK\label{aff4}
\and
Institut de Ciencies de l'Espai (ICE, CSIC), Campus UAB, Carrer de Can Magrans, s/n, 08193 Bellaterra (Cerdanyola del Valles), Spain\label{aff5}
\and
ESAC/ESA, Camino Bajo del Castillo, s/n., Urb. Villafranca del Castillo, 28692 Villanueva de la Ca\~nada, Madrid, Spain\label{aff6}
\and
INAF-Osservatorio Astronomico di Brera, Via Brera 28, 20122 Milano, Italy\label{aff7}
\and
INAF-Osservatorio di Astrofisica e Scienza dello Spazio di Bologna, Via Piero Gobetti 93/3, 40129 Bologna, Italy\label{aff8}
\and
IFPU, Institute for Fundamental Physics of the Universe, via Beirut 2, 34151 Trieste, Italy\label{aff9}
\and
INAF-Osservatorio Astronomico di Trieste, Via G. B. Tiepolo 11, 34143 Trieste, Italy\label{aff10}
\and
INFN, Sezione di Trieste, Via Valerio 2, 34127 Trieste TS, Italy\label{aff11}
\and
SISSA, International School for Advanced Studies, Via Bonomea 265, 34136 Trieste TS, Italy\label{aff12}
\and
Dipartimento di Fisica e Astronomia, Universit\`a di Bologna, Via Gobetti 93/2, 40129 Bologna, Italy\label{aff13}
\and
INFN-Sezione di Bologna, Viale Berti Pichat 6/2, 40127 Bologna, Italy\label{aff14}
\and
Institut de Physique Th\'eorique, CEA, CNRS, Universit\'e Paris-Saclay 91191 Gif-sur-Yvette Cedex, France\label{aff15}
\and
Institut d'Astrophysique de Paris, UMR 7095, CNRS, and Sorbonne Universit\'e, 98 bis boulevard Arago, 75014 Paris, France\label{aff16}
\and
Dipartimento di Fisica, Universit\`a di Genova, Via Dodecaneso 33, 16146, Genova, Italy\label{aff17}
\and
INFN-Sezione di Genova, Via Dodecaneso 33, 16146, Genova, Italy\label{aff18}
\and
Department of Physics "E. Pancini", University Federico II, Via Cinthia 6, 80126, Napoli, Italy\label{aff19}
\and
INAF-Osservatorio Astronomico di Capodimonte, Via Moiariello 16, 80131 Napoli, Italy\label{aff20}
\and
Dipartimento di Fisica, Universit\`a degli Studi di Torino, Via P. Giuria 1, 10125 Torino, Italy\label{aff21}
\and
INFN-Sezione di Torino, Via P. Giuria 1, 10125 Torino, Italy\label{aff22}
\and
INAF-Osservatorio Astrofisico di Torino, Via Osservatorio 20, 10025 Pino Torinese (TO), Italy\label{aff23}
\and
Leiden Observatory, Leiden University, Einsteinweg 55, 2333 CC Leiden, The Netherlands\label{aff24}
\and
INAF-IASF Milano, Via Alfonso Corti 12, 20133 Milano, Italy\label{aff25}
\and
Centro de Investigaciones Energ\'eticas, Medioambientales y Tecnol\'ogicas (CIEMAT), Avenida Complutense 40, 28040 Madrid, Spain\label{aff26}
\and
Port d'Informaci\'{o} Cient\'{i}fica, Campus UAB, C. Albareda s/n, 08193 Bellaterra (Barcelona), Spain\label{aff27}
\and
Institute of Space Sciences (ICE, CSIC), Campus UAB, Carrer de Can Magrans, s/n, 08193 Barcelona, Spain\label{aff28}
\and
Institut d'Estudis Espacials de Catalunya (IEEC),  Edifici RDIT, Campus UPC, 08860 Castelldefels, Barcelona, Spain\label{aff29}
\and
INAF-Osservatorio Astronomico di Roma, Via Frascati 33, 00078 Monteporzio Catone, Italy\label{aff30}
\and
INFN -- Sezione di Napoli, Via Cinthia 6, 80126, Napoli, Italy\label{aff31}
\and
Dipartimento di Fisica e Astronomia "Augusto Righi" - Alma Mater Studiorum Universit\`a di Bologna, Viale Berti Pichat 6/2, 40127 Bologna, Italy\label{aff32}
\and
Instituto de Astrof\'{\i}sica de Canarias, E-38205 La Laguna, Tenerife, Spain\label{aff33}
\and
Institute for Astronomy, University of Edinburgh, Royal Observatory, Blackford Hill, Edinburgh EH9 3HJ, UK\label{aff34}
\and
European Space Agency/ESRIN, Largo Galileo Galilei 1, 00044 Frascati, Roma, Italy\label{aff35}
\and
Universit\'e Lyon 1, CNRS/IN2P3, IP2I Lyon, Villeurbanne, France\label{aff36}
\and
Institut de Ci\`{e}ncies del Cosmos (ICCUB), Universitat de Barcelona (IEEC-UB), Mart\'{i} i Franqu\`{e}s 1, 08028 Barcelona, Spain\label{aff37}
\and
Instituci\'o Catalana de Recerca i Estudis Avan\c{c}ats (ICREA), Passeig de Llu\'{\i}s Companys 23, 08010 Barcelona, Spain\label{aff38}
\and
Institut de Ciencies de l'Espai (IEEC-CSIC), Campus UAB, Carrer de Can Magrans, s/n Cerdanyola del Vall\'es, 08193 Barcelona, Spain\label{aff39}
\and
Universit\'e Lyon 1, CNRS/IN2P3, IUF, IP2I Lyon, Villeurbanne, France \label{aff40}
\and
Departamento de F\'isica, Faculdade de Ci\^encias, Universidade de Lisboa, Edif\'icio C8, Campo Grande, PT1749-016 Lisboa, Portugal\label{aff41}
\and
Instituto de Astrof\'isica e Ci\^encias do Espa\c{c}o, Faculdade de Ci\^encias, Universidade de Lisboa, Campo Grande, 1749-016 Lisboa, Portugal\label{aff42}
\and
Department of Astronomy, University of Geneva, ch. d'Ecogia 16, 1290 Versoix, Switzerland\label{aff43}
\and
Universit\'e Paris-Saclay, CNRS, Institut d'astrophysique spatiale, 91405, Orsay, France\label{aff44}
\and
Aix-Marseille Universit\'e, CNRS/IN2P3, CPPM, Marseille, France\label{aff45}
\and
INAF-Istituto di Astrofisica e Planetologia Spaziali, via del Fosso del Cavaliere, 100, 00100 Roma, Italy\label{aff46}
\and
Universit\'e Paris-Saclay, Universit\'e Paris Cit\'e, CEA, CNRS, AIM, 91191, Gif-sur-Yvette, France\label{aff47}
\and
INFN-Bologna, Via Irnerio 46, 40126 Bologna, Italy\label{aff48}
\and
Department of Physics, University of Oxford, Keble Road, Oxford OX1 3RH, UK\label{aff49}
\and
University Observatory, LMU Faculty of Physics, Scheinerstr.~1, 81679 Munich, Germany\label{aff50}
\and
Max Planck Institute for Extraterrestrial Physics, Giessenbachstr. 1, 85748 Garching, Germany\label{aff51}
\and
INAF-Osservatorio Astronomico di Padova, Via dell'Osservatorio 5, 35122 Padova, Italy\label{aff52}
\and
Universit\"ats-Sternwarte M\"unchen, Fakult\"at f\"ur Physik, Ludwig-Maximilians-Universit\"at M\"unchen, Scheinerstr.~1, 81679 M\"unchen, Germany\label{aff53}
\and
Institute of Theoretical Astrophysics, University of Oslo, P.O. Box 1029 Blindern, 0315 Oslo, Norway\label{aff54}
\and
Jet Propulsion Laboratory, California Institute of Technology, 4800 Oak Grove Drive, Pasadena, CA, 91109, USA\label{aff55}
\and
Felix Hormuth Engineering, Goethestr. 17, 69181 Leimen, Germany\label{aff56}
\and
Technical University of Denmark, Elektrovej 327, 2800 Kgs. Lyngby, Denmark\label{aff57}
\and
Cosmic Dawn Center (DAWN), Denmark\label{aff58}
\and
Max-Planck-Institut f\"ur Astronomie, K\"onigstuhl 17, 69117 Heidelberg, Germany\label{aff59}
\and
NASA Goddard Space Flight Center, Greenbelt, MD 20771, USA\label{aff60}
\and
Universit\'e de Gen\`eve, D\'epartement de Physique Th\'eorique and Centre for Astroparticle Physics, 24 quai Ernest-Ansermet, CH-1211 Gen\`eve 4, Switzerland\label{aff61}
\and
Department of Physics, P.O. Box 64, University of Helsinki, 00014 Helsinki, Finland\label{aff62}
\and
Helsinki Institute of Physics, Gustaf H{\"a}llstr{\"o}min katu 2, University of Helsinki, 00014 Helsinki, Finland\label{aff63}
\and
Laboratoire d'etude de l'Univers et des phenomenes eXtremes, Observatoire de Paris, Universit\'e PSL, Sorbonne Universit\'e, CNRS, 92190 Meudon, France\label{aff64}
\and
SKAO, Jodrell Bank, Lower Withington, Macclesfield SK11 9FT, UK\label{aff65}
\and
Centre de Calcul de l'IN2P3/CNRS, 21 avenue Pierre de Coubertin 69627 Villeurbanne Cedex, France\label{aff66}
\and
Universit\"at Bonn, Argelander-Institut f\"ur Astronomie, Auf dem H\"ugel 71, 53121 Bonn, Germany\label{aff67}
\and
INFN-Sezione di Roma, Piazzale Aldo Moro, 2 - c/o Dipartimento di Fisica, Edificio G. Marconi, 00185 Roma, Italy\label{aff68}
\and
Aix-Marseille Universit\'e, CNRS, CNES, LAM, Marseille, France\label{aff69}
\and
Dipartimento di Fisica e Astronomia "Augusto Righi" - Alma Mater Studiorum Universit\`a di Bologna, via Piero Gobetti 93/2, 40129 Bologna, Italy\label{aff70}
\and
Department of Physics, Institute for Computational Cosmology, Durham University, South Road, Durham, DH1 3LE, UK\label{aff71}
\and
Universit\'e Paris Cit\'e, CNRS, Astroparticule et Cosmologie, 75013 Paris, France\label{aff72}
\and
CNRS-UCB International Research Laboratory, Centre Pierre Bin\'etruy, IRL2007, CPB-IN2P3, Berkeley, USA\label{aff73}
\and
Institute of Physics, Laboratory of Astrophysics, Ecole Polytechnique F\'ed\'erale de Lausanne (EPFL), Observatoire de Sauverny, 1290 Versoix, Switzerland\label{aff74}
\and
Dipartimento di Fisica - Sezione di Astronomia, Universit\`a di Trieste, Via Tiepolo 11, 34131 Trieste, Italy\label{aff75}
\and
Telespazio UK S.L. for European Space Agency (ESA), Camino bajo del Castillo, s/n, Urbanizacion Villafranca del Castillo, Villanueva de la Ca\~nada, 28692 Madrid, Spain\label{aff76}
\and
Institut de F\'{i}sica d'Altes Energies (IFAE), The Barcelona Institute of Science and Technology, Campus UAB, 08193 Bellaterra (Barcelona), Spain\label{aff77}
\and
European Space Agency/ESTEC, Keplerlaan 1, 2201 AZ Noordwijk, The Netherlands\label{aff78}
\and
Space Science Data Center, Italian Space Agency, via del Politecnico snc, 00133 Roma, Italy\label{aff79}
\and
Centre National d'Etudes Spatiales -- Centre spatial de Toulouse, 18 avenue Edouard Belin, 31401 Toulouse Cedex 9, France\label{aff80}
\and
Institute of Space Science, Str. Atomistilor, nr. 409 M\u{a}gurele, Ilfov, 077125, Romania\label{aff81}
\and
Dipartimento di Fisica e Astronomia "G. Galilei", Universit\`a di Padova, Via Marzolo 8, 35131 Padova, Italy\label{aff82}
\and
INFN-Padova, Via Marzolo 8, 35131 Padova, Italy\label{aff83}
\and
Instituto de F\'isica Te\'orica UAM-CSIC, Campus de Cantoblanco, 28049 Madrid, Spain\label{aff84}
\and
Institut de Recherche en Astrophysique et Plan\'etologie (IRAP), Universit\'e de Toulouse, CNRS, UPS, CNES, 14 Av. Edouard Belin, 31400 Toulouse, France\label{aff85}
\and
Universit\'e St Joseph; Faculty of Sciences, Beirut, Lebanon\label{aff86}
\and
Departamento de F\'isica, FCFM, Universidad de Chile, Blanco Encalada 2008, Santiago, Chile\label{aff87}
\and
Department of Physics and Helsinki Institute of Physics, Gustaf H\"allstr\"omin katu 2, University of Helsinki, 00014 Helsinki, Finland\label{aff88}
\and
Department of Physics, Royal Holloway, University of London, Surrey TW20 0EX, UK\label{aff89}
\and
Instituto de Astrof\'isica e Ci\^encias do Espa\c{c}o, Faculdade de Ci\^encias, Universidade de Lisboa, Tapada da Ajuda, 1349-018 Lisboa, Portugal\label{aff90}
\and
Cosmic Dawn Center (DAWN)\label{aff91}
\and
Niels Bohr Institute, University of Copenhagen, Jagtvej 128, 2200 Copenhagen, Denmark\label{aff92}
\and
Universidad Polit\'ecnica de Cartagena, Departamento de Electr\'onica y Tecnolog\'ia de Computadoras,  Plaza del Hospital 1, 30202 Cartagena, Spain\label{aff93}
\and
European University of Technology EUt+, European Union\label{aff94}
\and
Caltech/IPAC, 1200 E. California Blvd., Pasadena, CA 91125, USA\label{aff95}
\and
Dipartimento di Fisica e Scienze della Terra, Universit\`a degli Studi di Ferrara, Via Giuseppe Saragat 1, 44122 Ferrara, Italy\label{aff96}
\and
Istituto Nazionale di Fisica Nucleare, Sezione di Ferrara, Via Giuseppe Saragat 1, 44122 Ferrara, Italy\label{aff97}
\and
ICSC - Centro Nazionale di Ricerca in High Performance Computing, Big Data e Quantum Computing, Via Magnanelli 2, Bologna, Italy\label{aff98}
\and
INAF, Istituto di Radioastronomia, Via Piero Gobetti 101, 40129 Bologna, Italy\label{aff99}
\and
Astronomical Observatory of the Autonomous Region of the Aosta Valley (OAVdA), Loc. Lignan 39, I-11020, Nus (Aosta Valley), Italy\label{aff100}
\and
Universit\'e C\^{o}te d'Azur, Observatoire de la C\^{o}te d'Azur, CNRS, Laboratoire Lagrange, Bd de l'Observatoire, CS 34229, 06304 Nice cedex 4, France\label{aff101}
\and
Department of Mathematical Physics, Institute of Physics, University of S\~ao Paulo, R. do Mat\~ao 1371, 05508-090, S\~ao Paulo, SP, Brazil\label{aff102}
\and
CEA Saclay, DFR/IRFU, Service d'Astrophysique, Bat. 709, 91191 Gif-sur-Yvette, France\label{aff103}
\and
Universidad de La Laguna, Dpto. Astrof\'\i sica, E-38206 La Laguna, Tenerife, Spain\label{aff104}
\and
Oskar Klein Centre for Cosmoparticle Physics, Department of Physics, Stockholm University, Stockholm, SE-106 91, Sweden\label{aff105}
\and
Astrophysics Group, Blackett Laboratory, Imperial College London, London SW7 2AZ, UK\label{aff106}
\and
Univ. Grenoble Alpes, CNRS, Grenoble INP, LPSC-IN2P3, 53, Avenue des Martyrs, 38000 Grenoble, France\label{aff107}
\and
Dipartimento di Fisica, Sapienza Universit\`a di Roma, Piazzale Aldo Moro 2, 00185 Roma, Italy\label{aff108}
\and
Institut d'Astrophysique de Paris, 98bis Boulevard Arago, 75014, Paris, France\label{aff109}
\and
ICL, Junia, Universit\'e Catholique de Lille, LITL, 59000 Lille, France\label{aff110}
\and
Center for Astrophysics and Cosmology, University of Nova Gorica, Nova Gorica, Slovenia\label{aff111}
\and
CERCA/ISO, Department of Physics, Case Western Reserve University, 10900 Euclid Avenue, Cleveland, OH 44106, USA\label{aff112}
\and
Laboratoire Univers et Th\'eorie, Observatoire de Paris, Universit\'e PSL, Universit\'e Paris Cit\'e, CNRS, 92190 Meudon, France\label{aff113}
\and
Dipartimento di Fisica "Aldo Pontremoli", Universit\`a degli Studi di Milano, Via Celoria 16, 20133 Milano, Italy\label{aff114}
\and
INFN-Sezione di Milano, Via Celoria 16, 20133 Milano, Italy\label{aff115}
\and
Departamento de F{\'\i}sica Fundamental. Universidad de Salamanca. Plaza de la Merced s/n. 37008 Salamanca, Spain\label{aff116}
\and
Universiteit Gent, Department of Physics and Astronomy, Proeftuinstraat 86 N3, 9000 Ghent, Belgium
\label{aff117}
\and
Universit\'e de Strasbourg, CNRS, Observatoire astronomique de Strasbourg, UMR 7550, 67000 Strasbourg, France\label{aff118}
\and
Center for Data-Driven Discovery, Kavli IPMU (WPI), UTIAS, The University of Tokyo, Kashiwa, Chiba 277-8583, Japan\label{aff119}
\and
Waterloo Centre for Astrophysics, University of Waterloo, Waterloo, Ontario N2L 3G1, Canada\label{aff120}
\and
Scuola Superiore Meridionale, Via Mezzocannone 4, 80138, Napoli, Italy\label{aff121}
\and
Jodrell Bank Centre for Astrophysics, Department of Physics and Astronomy, University of Manchester, Oxford Road, Manchester M13 9PL, UK\label{aff122}
\and
California Institute of Technology, 1200 E California Blvd, Pasadena, CA 91125, USA\label{aff123}
\and
Department of Physics, Astronomy and Mathematics, University of Hertfordshire, College Lane, Hatfield AL10 9AB, UK\label{aff124}
\and
Institute of Cosmology and Gravitation, University of Portsmouth, Portsmouth PO1 3FX, UK\label{aff125}
\and
Department of Physics \& Astronomy, University of California Irvine, Irvine CA 92697, USA\label{aff126}
\and
Kapteyn Astronomical Institute, University of Groningen, PO Box 800, 9700 AV Groningen, The Netherlands\label{aff127}
\and
Universit\"at Innsbruck, Institut f\"ur Astro- und Teilchenphysik, Technikerstr. 25/8, 6020 Innsbruck, Austria\label{aff128}
\and
Institute for Astronomy, University of Hawaii, 2680 Woodlawn Drive, Honolulu, HI 96822, USA\label{aff129}
\and
Departamento F\'isica Aplicada, Universidad Polit\'ecnica de Cartagena, Campus Muralla del Mar, 30202 Cartagena, Murcia, Spain\label{aff130}
\and
Instituto de F\'isica de Cantabria, Edificio Juan Jord\'a, Avenida de los Castros, 39005 Santander, Spain\label{aff131}
\and
Department of Mathematics and Physics E. De Giorgi, University of Salento, Via per Arnesano, CP-I93, 73100, Lecce, Italy\label{aff132}
\and
INFN, Sezione di Lecce, Via per Arnesano, CP-193, 73100, Lecce, Italy\label{aff133}
\and
INAF-Sezione di Lecce, c/o Dipartimento Matematica e Fisica, Via per Arnesano, 73100, Lecce, Italy\label{aff134}
\and
Department of Computer Science, Aalto University, PO Box 15400, Espoo, FI-00 076, Finland\label{aff135}
\and
Ruhr University Bochum, Faculty of Physics and Astronomy, Astronomical Institute (AIRUB), German Centre for Cosmological Lensing (GCCL), 44780 Bochum, Germany\label{aff136}
\and
DARK, Niels Bohr Institute, University of Copenhagen, Jagtvej 155, 2200 Copenhagen, Denmark\label{aff137}
\and
Department of Physics, Centre for Extragalactic Astronomy, Durham University, South Road, Durham, DH1 3LE, UK\label{aff138}
\and
Astrophysics Research Centre, University of KwaZulu-Natal, Westville Campus, Durban 4041, South Africa\label{aff139}
\and
School of Mathematics, Statistics \& Computer Science, University of KwaZulu-Natal, Westville Campus, Durban 4041, South Africa\label{aff140}
\and
Department of Physics and Astronomy, Vesilinnantie 5, University of Turku, 20014 Turku, Finland\label{aff141}
\and
Finnish Centre for Astronomy with ESO (FINCA), Quantum, Vesilinnantie 5, University of Turku, 20014 Turku, Finland\label{aff142}
\and
Serco for European Space Agency (ESA), Camino bajo del Castillo, s/n, Urbanizacion Villafranca del Castillo, Villanueva de la Ca\~nada, 28692 Madrid, Spain\label{aff143}
\and
ARC Centre of Excellence for Dark Matter Particle Physics, Melbourne, Australia\label{aff144}
\and
Centre for Astrophysics \& Supercomputing, Swinburne University of Technology,  Hawthorn, Victoria 3122, Australia\label{aff145}
\and
Department of Physics and Astronomy, University of the Western Cape, Bellville, Cape Town, 7535, South Africa\label{aff146}
\and
MTA-CSFK Lend\"ulet Large-Scale Structure Research Group, Konkoly-Thege Mikl\'os \'ut 15-17, H-1121 Budapest, Hungary\label{aff147}
\and
Konkoly Observatory, HUN-REN CSFK, MTA Centre of Excellence, Budapest, Konkoly Thege Mikl\'os {\'u}t 15-17. H-1121, Hungary\label{aff148}
\and
Ohio University, Physics \& Astronomy Department,1 Ohio University, Athens, OH 45701, USA\label{aff149}
\and
Brazilian Center for Research in Physics (CBPF), Dr. Xavier Sigaud st. 150, zip 22290-180, Rio de Janeiro, RJ, Brazil\label{aff150}
\and
Institute for Theoretical Particle Physics and Cosmology (TTK), RWTH Aachen University, 52056 Aachen, Germany\label{aff151}
\and
INAF-Osservatorio Astrofisico di Arcetri, Largo E. Fermi 5, 50125, Firenze, Italy\label{aff152}
\and
Centro de Astrof\'{\i}sica da Universidade do Porto, Rua das Estrelas, 4150-762 Porto, Portugal\label{aff153}
\and
Instituto de Astrof\'isica e Ci\^encias do Espa\c{c}o, Universidade do Porto, CAUP, Rua das Estrelas, PT4150-762 Porto, Portugal\label{aff154}
\and
Dipartimento di Fisica, Universit\`a di Roma Tor Vergata, Via della Ricerca Scientifica 1, Roma, Italy\label{aff155}
\and
INFN, Sezione di Roma 2, Via della Ricerca Scientifica 1, Roma, Italy\label{aff156}
\and
HE Space for European Space Agency (ESA), Camino bajo del Castillo, s/n, Urbanizacion Villafranca del Castillo, Villanueva de la Ca\~nada, 28692 Madrid, Spain\label{aff157}
\and
Dipartimento di Fisica ``G. Occhialini", Universit\`a degli Studi di Milano Bicocca, Piazza della Scienza 3, 20126 Milano, Italy\label{aff158}
\and
Department of Astrophysics, University of Zurich, Winterthurerstrasse 190, 8057 Zurich, Switzerland\label{aff159}
\and
Theoretical astrophysics, Department of Physics and Astronomy, Uppsala University, Box 524, 751 20 Uppsala, Sweden\label{aff160}
\and
Mathematical Institute, University of Leiden, Einsteinweg 55, 2333 CA Leiden, The Netherlands\label{aff161}
\and
Institute of Astronomy, University of Cambridge, Madingley Road, Cambridge CB3 0HA, UK\label{aff162}
\and
Kobayashi-Maskawa Institute for the Origin of Particles and the Universe, Nagoya University, Chikusa-ku, Nagoya, 464-8602, Japan\label{aff163}
\and
Institute for Particle Physics and Astrophysics, Dept. of Physics, ETH Zurich, Wolfgang-Pauli-Strasse 27, 8093 Zurich, Switzerland\label{aff164}
\and
Department of Astrophysical Sciences, Peyton Hall, Princeton University, Princeton, NJ 08544, USA\label{aff165}
\and
Fakult\"at f\"ur Physik, Universit\"at Bielefeld, Postfach 100131, 33501 Bielefeld, Germany\label{aff166}
\and
School of Mathematics, Statistics and Physics, Newcastle University, Herschel Building, Newcastle-upon-Tyne, NE1 7RU, UK\label{aff167}
\and
SRON Netherlands Institute for Space Research, Landleven 12, 9747 AD, Groningen, The Netherlands\label{aff168}}    

 \abstract{
 We present \texttt{NaturalSpice}, a new \texttt{Python} implementation of the \texttt{PolSpice} angular power spectrum estimator. We show that the mixing matrix translating between pseudo and full-sky power spectra has eigenvalues given by the correlation function of the survey mask. Therefore, harmonic and real-space approaches to modelling mask effects are equivalent in the absence of a band limit. However, in real space the regularisation required for a given footprint can be consistently derived from its correlation function, which \texttt{NaturalSpice} employs by reformulating mode unmixing as a pseudo-inverse problem. We benchmark \texttt{NaturalSpice} against the widely used \texttt{NaMaster} algorithm using 200 fixed-cosmology light cone simulations applying three footprints representative of Euclid DR1 analyses ($\sim 1600$, $\sim 400$, and $\sim 20\sqdeg$). With a logistic apodisation function, \texttt{NaturalSpice} returns unbiased estimates with approximately 15, 30, and 60 per cent lower variance than \texttt{NaMaster} for the large, medium, and small footprints, respectively, and less correlated covariances for the two smaller footprints. These gains are concentrated at $\ell < 100$, while the two algorithms have equivalent performance on smaller scales. \texttt{NaturalSpice} is therefore particularly suited to early Stage-IV analyses with small-area but high-density datasets and as a tool to check for large-scale systematic effects. It is publicly available as part of the \Euclid angular power spectra code \texttt{Heracles}.
 }
%
% Provide up to five key words:
%

\titlerunning{\textit{Euclid} NaturalSpice}
\authorrunning{Euclid Collaboration: J. Ruiz-Zapatero et al.}

\maketitle
   
\section{Introduction} \label{Sect: introduction}
 
The presence of masks in angular power spectra analyses induces an overall reduction in the amplitude of the signal that manifests as power being mixed between different modes \citep{Peebles_Cls}. Thus, the impact of masking is often referred to as `mixing' or `coupling'.  Mixing is particularly consequential when analysing polarised fields, where power is also shifted across the polarised components -- a phenomenon known as `leakage'. It is often desired to undo the impact of masking in the observed angular power spectra, primarily for two reasons. First, the theoretical expectation of the angular power spectra is naturally computed on the full sky. Second, the presence of systematic effects can be obfuscated by the impact of masking on the signal. As a result, the cosmology community has proposed various algorithms to `unmix' or `decouple' the observed angular power spectra. On the one hand, the \texttt{MASTER} algorithm \citep{0105302} tackles the problem in harmonic space; on the other hand, the \texttt{PolSpice} algorithm \citep{10.1086/319105, 0303414, 2204.13721} tackles it in real space.
 
Cosmic Microwave Background (CMB) analyses typically mask relatively small portions of the sky where the Galactic plane and other sources of contamination lie in an otherwise full-sky map. Galaxy surveys are fundamentally different: only small sections of the sky are observed, introducing a maximum angular separation between the pairs of sources that can be found. This maximum angular separation, after which our observations vanish, makes the impact of the mask not directly invertible. \texttt{MASTER} and \texttt{PolSpice} estimate the full-sky angular power spectra by regularising the mask's impact. The \texttt{MASTER} methodology has been widely implemented in galaxy surveys to great success \citep{HSC_DR3, 2503.19441, DESY6, Unions}; see the \texttt{NaMaster} library by \citet{1809.09603}. Similarly, \texttt{PolSpice} has been primarily used in CMB analyses \citep{planck13_lkl, planck15_lkl, planck18_lkl, 1512.04535} even when only a small portion of the sky was observed \citep{BICEP}. Nonetheless, the regularisation of each algorithm plays an outsized role in determining the bias and variance of the final estimate as the size of the footprint decreases. This raises the question of whether these regularisation schemes are `optimal'.
 
With the advent of Stage-IV surveys such as the Vera C. Rubin Observatory's Legacy Survey of Space and Time \citep[LSST;][]{LSST} or the Euclid Wide Survey \citep{Euclid}, it is paramount to provide a systematic resolution to these issues both to extract as much information as possible from the upcoming data sets as well as to identify and mitigate potential systematic effects. Similarly, the ability of future CMB experiments, such as the Simons Observatory~\citep{Simons_observatory}, to detect primordial gravitational waves will hinge on the accurate reconstruction of the large signals of CMB polarisation. This is particularly important for their early data releases for which only small fractions of the sky will be observed -- but with an unprecedented level of precision -- making all the problems listed above more pressing. 

In this paper we reformulate unmixing as a diagonalisation problem. Within this framework, undoing the impact of masking amounts to inverting the mixing matrix associated with a given footprint. We show that the eigenvalues of the mixing matrix are given by the correlation function of the footprint. In doing so, we show that harmonic and real space strategies to undo the impact of mixing are equivalent in the limit where infinite harmonic modes are considered. However, in the presence of a band limit, we show that working directly in real space offers a series of advantages. The most important of these is that the amount of regularisation required by the footprint to become invertible can be directly derived from the correlation function of the mask. 

Armed with this knowledge, we develop our own implementation of the \texttt{PolSpice} algorithm, named \texttt{NaturalSpice}. To test and benchmark \texttt{NaturalSpice}, we use a subset of the Gower Street \citep[GS;][]{GowerSt} simulations at fixed cosmology. By measuring the full-sky angular power spectra of the simulations and comparing these to the spectra reconstructed by the different algorithms we can assess the effectiveness of each method in the presence of different masks. Our results show that, with adequate regularisation, \texttt{NaturalSpice} yields higher signal-to-noise ratio (S/N) estimates of angular power spectra than the state-of-the-art code \texttt{NaMaster}. \texttt{NaturalSpice} achieves this while remaining unbiased and producing invertible, less correlated, covariance matrices. In addition to this, \texttt{NaturalSpice} is also fast due to outsourcing expensive computations to \texttt{C} libraries and not relying on any optimisation or marginalisation process.
 
The paper is structured as follows. In Sect.~\ref{sect: observations} we discuss the angular power spectrum formalism for the different types of galaxy survey observations. In Sect.~\ref{sect: unmixing} we formally present \texttt{NaturalSpice}; this is done alongside the current unmixing algorithms in order to showcase their differences. In Sect.~\ref{Sect: data} we describe the data we will use to test the algorithms, while in Sect.~\ref{Sect: Results} we compare the various algorithms for different types of masks. Finally we give our conclusions in Sect.~\ref{Sect: Conclusions}.
 
\section{Angular power spectra} \label{sect: observations}
\subsection{Full sky}
Galaxy surveys observe various properties of luminous galaxies. These properties are described by discrete spherical functions understood to be discrete samples of a continuous field, $f(\nv)$, where $\nv$ is the unit vector on the (celestial) sphere. These properties can be classified in terms of their spin-weight; a function~$f(\nv)$ defined on the sphere has a spin-weight $s$ if the function value $f(\nv)$ transforms under a rotation~$\alpha$ of the tangent plane as
\begin{equation}
\label{eq:spinweight}
    f(\nv) \mapsto \mathrm{e}^{-\I s \alpha} \, f(\nv) \;.
\end{equation}
Scalar fields such as number density or the CMB temperature are of spin-weight $s=0$ (i.e. they are invariant under tangent plane rotations) while cosmic shear or CMB polarisation have spin-weight $s=2$ (i.e. they are invariant under tangent plane rotations of $\pi$ radians).
 
The goal of astronomical analyses is to learn the statistical properties of these fields. This task reduces to that of learning the moments of their distribution, $\langle f^{n}(\nv) \rangle$, where $n$ denotes the moment order. For this it is useful to expand in terms of spin-weighted spherical harmonics \citep[SWSH; for details see for example][]{castro2005},
\begin{equation}
    f(\nv) = \sum_{\ell m} f_{\ell m} \, {}_sY_{\ell m}(\nv) \;.
\end{equation}
Here $_sY_{\ell m}$ is a spherical harmonic with spin-weight $s$, and 
\begin{equation}
\label{eq:def-f-ylm}
    f_{\ell m} = \int \! f(\nv) \, [_sY_{\ell m}(\nv)]^* \, \diff^2\nv \,.
\end{equation}

Often these fields are defined as over-densities over a mean field, meaning that their first moment is zero by definition. Moreover, if the fields are Gaussian-distributed, the second moment contains all the field information. Therefore, traditionally most efforts have been put into estimating the second moment. 

Let $f$ (resp. $f'$) be a spherical function of spin-weight $s$ (resp. $s'$). We define the angular (cross-)power spectrum to be an average over the $m$-modes of products of harmonic coefficients, 
\begin{equation} \label{eq: cl_estimator}
    C_{\ell}^{f\!f'} \equiv \frac{1}{2\ell+1} \sum_m f_{\ell m}^* f'_{\ell m} \,.
\end{equation}
Here $f_{\ell m}^*$ means $(f_{\ell m})^*$. The associated real-space correlation function is defined by
\begin{equation}
\label{eq:ev-cf-cl}
      \fullxi{}(\theta) \equiv \sum_{\ell}\frac{2\ell+1}{4\pi}  C_\ell^{f\!f'}  \, d^\ell_{ss'}(\theta) \,,
\end{equation}
where $d^\ell_{ss'}(\theta)$ is the Wigner $d$-matrix. Inverting Eq.~\eqref{eq:ev-cf-cl} via the $d$-matrix orthogonality relation yields
\begin{equation}
\label{eq:ev-cl-cf}
     C_\ell^{f \!f'} \
    = 2\pi \int_{0}^{\pi} \!
        \fullxi{}(\theta)  \, d^\ell_{ss'}(\theta) \, \sin\theta \, \diff\theta \,,
\end{equation}
which shows the equivalence of both representations (in the absence of band-limits).

These empirical definitions of power spectrum and correlation function are used as they are applicable to arbitrary spin-weight functions. In the case that $f$ and $f'$ are samples from homogeneous random fields, the covariance of the harmonic coefficients is diagonal and, via the ergodic principle, matches the expectation of the empirically defined power spectrum of Eq.~\eqref{eq: cl_estimator},
\begin{equation}
\label{eq:ev-flm-prod}
    \langle f^*_{\ell m} \, f'_{\ell'm'} \rangle
    = \delta^{\rm K}_{\ell\ell'} \, \delta^{\rm K}_{mm'} \,  \langle C_\ell^{f\!f'} \rangle \;.
\end{equation}
Here $\delta^{\rm  K}$ is the Kronecker delta function, and the expectation values are taken over realisations of the random fields. Similarly, the SWSH addition theorem shows that in this case the expectation of the real-space correlation function is equal to the expectation of the product of $f^*$ and $f'$ at separated points (the latter being the more usual definition of correlation function).  In the above we write `homogeneous' to mean `statistically invariant under rotations' and we note that `homogeneous' and `isotropic' are equivalent on the sphere.

Finally, we explicitly consider spin-weight 2 fields as they are relevant to cosmic shear. The associated SWSH expansion of the Stokes vectors of a field $\gamma$ of spin-weight 2 is
\begin{equation} \label{eq:EB}
 \gamma(\nv) = \sum_{\ell m} (E_{\ell m} + \I B_{\ell m}) \, _{2} Y_{\ell m}(\nv) \; ,
\end{equation}
where $E_{\ell m }$ and $B_{\ell m}$ are the gradient and curl components of the tensor field \citep{Lewis2005}. The sign convention of Eq.~\eqref{eq:EB} follows \citet{Heracles} and is opposite to the \texttt{HEALPix} convention \citep{HealPix}. Inserting Eq. \eqref{eq:EB} into Eq. \eqref{eq: cl_estimator} we obtain the angular power spectra of the field
\begin{equation} \label{eq: cl_gg}
C_\ell^{\gamma \!\gamma'} = C_\ell^{E\!E'} + C_\ell^{B\!B'} + \I (C_\ell^{E\!B'} - C_\ell^{B\!E'})\, .
\end{equation}
In real space Eq.~\eqref{eq: cl_gg} becomes a complex correlation function with a real part given by 
\begin{equation} \label{eq: xi_gg}
\Re[\xi^{\gamma \!\gamma'}(\theta)] = \sum_\ell \frac{2\ell+1}{4\pi} (C_\ell^{E\!E'} + C_\ell^{B\!B'}) \, d^\ell_{22}(\theta) \, ,
\end{equation}
and an imaginary part given by
\begin{equation} \label{eq: im_xi_gg}
\Im[\xi^{\gamma \!\gamma'}(\theta)] = \sum_\ell \frac{2\ell+1}{4\pi} (C_\ell^{E\!B'} - C_\ell^{B\!E'}) \, d^\ell_{22}(\theta) \, .
\end{equation}
This shows that in real space the information from the $E$- and $B$-components of the shear field are mixed. To allow us to isolate the two contributions using real-space measurements, we consider also the pseudo-correlation function $\xi^{\gamma^* \!\gamma'}(\theta)$ given by 
\begin{equation} \label{eq: pseudo-xi_gg}
\Re[\xi^{\gamma^* \!\gamma'}(\theta)] = \sum_\ell \frac{2\ell+1}{4\pi} (C_\ell^{E\!E'} - C_\ell^{B\!B'}) \, d^\ell_{2-2}(\theta) \, ,
\end{equation}
and
\begin{equation} \label{eq: imag_pseudo-xi_gg}
\Im[\xi^{\gamma^* \!\gamma'}(\theta)] = \sum_\ell \frac{2\ell+1}{4\pi} (C_\ell^{E\!E'} + C_\ell^{B\!B'}) \, d^\ell_{2-2}(\theta) \, .
\end{equation}
Here we have used the symmetry $d^{\ell}_{-22}=d^{\ell}_{2-2}$. We denote these two correlation functions by $\xi^{\pm}$
\begin{equation} \label{eq: xi_pm}
\xi^{\pm}(\theta) \equiv \sum_\ell \frac{2\ell+1}{4\pi} (C_\ell^{E\!E'} \pm C_\ell^{B\!B'}) \, d^\ell_{2\pm2}(\theta) \, ,
\end{equation}
and
\begin{equation} \label{eq: xi_cross_pm}
\xi^{\times \pm}(\theta) \equiv \sum_\ell \frac{2\ell+1}{4\pi} (C_\ell^{E\!B'} \pm C_\ell^{B\!E'}) \, d^\ell_{2\mp2}(\theta) \, ,
\end{equation}
which between them contain enough information to separate the $E$- and $B$-modes.

\subsection{Partial sky}
The observed partial-sky function $\partfn{}$ is the product of a full-sky function $\fullfn$ (assumed to be a draw from a homogeneous distribution) and a real-valued deterministic, not necessarily isotropic weighting function $w$ that masks the sky,
\begin{equation} \label{eq:prod}
    \partfn{}(\nv) = \fullfn(\nv) \, w(\nv) \;.
\end{equation}
The expectation value of the angular correlation function of the observed masked field is the product of the expectation value of the correlation function of the underlying full-sky field and the correlation function of the weighting function, i.e.
\begin{equation}
\label{eq:ev-cf-prod}
    \langle \partxi{}(\theta) \rangle
    = \langle \fullxi{}(\theta) \rangle \, \xi^{w\!w'}(\theta) \;.
\end{equation}
We recall that $\xi^{w\!w'}$ is defined by applying Eqs.~\eqref{eq: cl_estimator} and \eqref{eq:ev-cf-cl} to the spherical harmonic coefficients of $w$ and $w'$. This is equivalent to taking the azimuthal average of the full correlation function of the anisotropic weighting field, which destroys the directional information encoded in it. Therefore, Eq.~\eqref{eq:ev-cf-prod} only holds in expectation since the statistics of the realised partial-sky spherical function is sensitive to rotations of the mask. However, taking the average over many realisations of the field removes this rotational asymmetry since it does not matter where one probes a homogeneous field \citep{Heracles}.

Thus, we see that in real space the correlation function of the mask is diagonal, meaning that the perceived mixing of power between harmonic modes is not an effect of the masking itself but rather of the change of basis from real space to harmonic space. Let $s_1$ and $s_2$ be the spin-weights of $\fullfn$ and $w$, respectively, so that $s=s_1+s_2$ is the spin-weight of the partial-sky field $\partfn{}$. In our particular case $s_2=0$ (as $w$ is a scalar weighting function), so the spin of the field is unchanged under masking. Combining Eqs.~\eqref{eq:ev-cl-cf}, \eqref{eq:ev-cf-prod}, and \eqref{eq:ev-cf-cl} yields
\begin{align}
    \langle \partC{\ell} \rangle &= 2\pi \sum_{\ell_1 \ell_2} \frac{2\ell_1+1}{4\pi}  \frac{2\ell_2+1}{4\pi} \, \langle \fullC{\ell_1} \rangle \, C^{w\!w'}_{\ell_2} \nonumber \\
    & \quad \times \int_{0}^{\pi} \! d^\ell_{ss'}(\theta) \, d^{\ell_1}_{s_1s_1'}(\theta) \, d^{\ell_2}_{s_2s_2'}(\theta) \, \sin\theta \, \diff\theta \nonumber \\
        &=  \sum_{\ell_1} M_{\ell\ell_1}^{w\!w'} \langle \fullC{\ell_1} \rangle \;.\label{eq:ev-cl-mm} 
\end{align}
Here we have defined a convolution kernel 
\begin{equation}
\label{eq:mm}
    M_{\ell\ell_1}^{ww'} = (-1)^{s-s'} \frac{2\ell_1+1}{4\pi} \sum_{\ell_2} \mathcal{K}_{\ell \ell_1 \ell_2} \, C_{\ell_2}^{ww'} \, ,
\end{equation}
known as the mixing or coupling matrix, where
\begin{equation}
\mathcal{K}_{\ell \ell_1 \ell_2} = (2\ell_2+1) \,
        \threej{\ell_1,s_1}{\ell_2,s_2}{\ell,-s} \,
        \threej{\ell_1,s_1'}{\ell_2,s_2'}{\ell,-s'} \, .
\end{equation}
In the above we have used the symmetry $d^\ell_{ss'} = (-1)^{s-s'}d^\ell_{-s,-s'}$ as well as the triple-$d$ integral (a reduced form of the Gaunt integral). Within this formalism, the mixing of modes due to masking follows naturally from transforming the masking operation from real space, where it is a diagonal operator, to a different basis, where it becomes a dense matrix.

In the case of fields of spin-weight 2, the change of basis not only shifts power within the $E$ and $B$ spectra but also across them (`$E$--$B$-mode leakage') since they are different components of a single complex field. For such fields, Eq.~\eqref{eq:ev-cl-mm} becomes
\begin{gather}
 \begin{bmatrix} 
 \langle \tilde{C}^{E\!E'} \rangle \\
 \langle \tilde{C}^{B\!B'} \rangle \\
 \langle \tilde{C}^{E\!B'} \rangle \\
 \langle \tilde{C}^{B\!E'} \rangle
 \end{bmatrix}_{\ell}
 = \nonumber \\
 \qquad \qquad \sum_{\ell_1}
\begin{bmatrix}
   M^{+} & M^{-} & 0 & 0 \\
   M^{-} & M^{+} & 0 & 0 \\
   0 & 0 & M^{+} & -M^{-} \\
   0 & 0 & -M^{-} & M^{+}
\end{bmatrix}_{\ell\ell_1}
\begin{bmatrix} 
    \langle C^{E\!E'} \rangle \\
    \langle C^{B\!B'} \rangle \\
    \langle C^{E\!B'} \rangle \\
    \langle C^{B\!E'} \rangle
\end{bmatrix}_{\ell_1} \; ,
\label{eq: M_spin2}
\end{gather}
where
\begin{equation} \label{eq: M_rot}
M_{\ell\ell_1}^{\pm} = (-1)^{s-s'} \frac{2\ell_1+1}{4\pi} \sum_{\ell_2} \mathcal{K}_{\ell \ell_1 \ell_2} \, \frac{1 \pm (-1)^{\ell_1 + \ell_2 + \ell}}{2} \, C_{\ell_2}^{ww'} \,.
\end{equation}
We see that $M^{-}$ controls the mixing between $E$- and $B$-modes. It is clear that Eq.~\eqref{eq: M_spin2} becomes block-diagonal under a simple rotation. This can be seen by introducing the variables
\begin{gather} 
    C_\ell^\pm = C_\ell^{EE'} \pm C_\ell^{BB'} \; , \label{eq: cl_rot} \\
    C_\ell^{\times \pm} = C_\ell^{EB'} \pm C_\ell^{BE'} \; ,  \label{eq: cl_cross_rot}
\end{gather}
such that 
\begin{gather}
    \langle \tilde{C}_{\ell}^\pm \rangle = \sum_{\ell_1} \left( M^{+} \pm M^{-} \right)_{\ell \ell_1} \langle C_{\ell_1}^\pm \rangle \; ,  \label{eq:master_rot} \\
    \langle \tilde{C}_{\ell}^{\times \pm} \rangle = \sum_{\ell_1} \left( M^{+} \mp M^{-} \right)_{\ell \ell_1} \langle C_{\ell_1}^{\times \pm} \rangle \; , \label{eq:master_cross_rot}
\end{gather}
which contain the same mixing of $E$- and $B$-mode information as their equivalent correlation functions, Eqs.~(\ref{eq: xi_pm}) and (\ref{eq: xi_cross_pm}), respectively.
 
Due to the presence of statistical and systematic noise, the estimation of the SWSH coefficients from real data is noisy. A popular way of reducing this noise is to average $\langle C_\ell^{f \!f'} \rangle$ over a mode-window function. This can be done by applying to both sides of Eq.~\eqref{eq:ev-cl-mm} a binning matrix with entries 
\begin{equation}
\label{eq:binning}
    B_{\ell q} = \begin{cases} (\ell_q - \ell_{q-1})^{-1} & \text{if } \ell_{q-1} \le \ell < \ell_{q} \\ 0 & \text{otherwise} \end{cases} \, ,
\end{equation}
for bin boundaries $\ell_{\text{min}} = \ell_0 < \ell_1 < \ldots < \ell_Q = \ell_{\text{max}}+1$.
 
\section{Unmixing algorithms} \label{sect: unmixing}
\subsection{The problem} \label{sect: NaturalSpice}
Our goal is to find estimators $\estimator{C}_\ell$ of the full-sky angular power spectra $C_\ell$. We want these estimators to be functions of the corresponding (observable) partial sky quantities $\tilde{C}_\ell$, and to be unbiased i.e. $\langle \estimator{C}_\ell \rangle = \langle C_\ell \rangle$. In particular, we seek an unbiased estimator $\estimator{C}^{f\!f'}_\ell$ of the full-sky angular power spectra. Any unbiased estimator $\estimator{C}_\ell$ must be expressible as a linear transformation of Eq.~\eqref{eq:ev-cl-mm}, so that
\begin{equation}
\label{eq:master}
     \estimator{C}_{\ell}^{ff'} \equiv  \sum_{\ell_1} A_{\ell\ell_1} \,  \tilde{C}_{\ell_1}^{f\!f'}
    =   \sum_{\ell_1\ell_2} A_{\ell\ell_1} \, M_{\ell_1\ell_2}^{w\!w'}  C_{\ell_2}^{f\!f'} \;.
\end{equation}
From Eq.~\eqref{eq:master} it follows that $A_{\ell \ell'}\equiv (M^{w\!w'})^{-1}_{\ell\ell'}$. 

Thus, the problem of angular power spectrum estimation fundamentally boils down to writing an expression for the inverse mixing matrix. In order to do so let us use the equivalent expression to Eq.~\eqref{eq:ev-cl-mm} in real space, Eq.~\eqref{eq:ev-cf-prod}. Moving $\xi^{ww'}(\theta)$ to the right-hand side of Eq.~\eqref{eq:ev-cf-prod} yields the estimator for the full-sky angular correlation function
\begin{equation} \label{eq:xi_hat}
         \estimator{\xi}^{f\!f'}(\theta) 
    = \left[\xi^{w\!w'}(\theta)\right]^{-1}  \partxi{}(\theta) \, ,
\end{equation}
which is unbiased by virtue of Eq.~\eqref{eq:ev-cl-mm}. Writing the correlation functions of the fields explicitly in terms of the measured angular power spectra and transforming back to harmonic space yields the expression
\begin{gather} 
        \estimator{C}^{f\!f'}_{\ell}   = 
          \! \sum_{\ell_1}\frac{2\ell_1+1}{2}  \tilde{C}_{\ell_1}^{f\!f'} \nonumber \\ \times \int_{0}^{\pi} \left[\xi^{ w\!w'}(\theta)\right]^{-1} 
            \! d^{\ell}_{ss'}(\theta) \, d^{\ell_1}_{ss'}(\theta) \, \sin\theta \, \diff\theta  \, ,\label{eq:naturalspice}
\end{gather}
which spells out the form of the inverse mixing matrix
\begin{equation} \label{eq: All}
    A_{\ell\ell'} = \frac{2\ell'+1}{2} \int^{\pi}_{0} \bigl[\xi^{ww'}(\theta)\bigr]^{-1} \, d^{\ell}_{ss'}(\theta) \, d^{\ell'}_{ss'}(\theta) \, \sin\theta \, \diff\theta \, .
\end{equation}

Obtaining an expression for the inverse mixing matrix of spin-2 fields is slightly more involved due to the mixing of $E$- and $B$-modes. However, the same strategy can be pursued by considering the rotated spectra in Eqs.~(\ref{eq: cl_rot}) and (\ref{eq: cl_cross_rot}). Then, one can write the associated estimators 
\begin{gather}
     \estimator{C}^{\pm}_{\ell} = \sum_{\ell'} A_{\ell \ell'}^{\pm} \tilde{C}_{\ell'}^{\pm} \; , \\
     \estimator{C}^{\times \pm}_{\ell} = \sum_{\ell'} A_{\ell \ell'}^{\mp} \tilde{C}_{\ell'}^{\times \pm} \;,
\end{gather}
where
\begin{equation} \label{eq: All_pm}
     A_{\ell \ell'}^{\pm} \equiv \frac{2\ell'+1}{2} \int_{0}^{\pi} \left[ \xi^{ww'}(\theta)\right]^{-1} d^{\ell}_{2\pm2}(\theta) \, d^{\ell'}_{2\pm2}(\theta) \, \sin\theta \, \diff\theta \, ,
\end{equation}
which satisfies
\begin{equation} \label{eq: AM_pm}
    \sum_{\ell_1} A^{\pm}_{\ell \ell_1} (M^{+} \pm M^{-})_{\ell_1 \ell_2}  = \delta^{\rm K}_{\ell \ell_2} \, .
\end{equation}
Estimates of the corresponding $E$-mode and $B$-mode auto- and cross-correlation spectra can be obtained by considering the linear combinations
\begin{gather}
     \estimator{C}^{EE'}_{\ell} =  \frac{1}{2} (\estimator{C}^{+}_{\ell} + \estimator{C}^{-}_{\ell}) \, , \\
     \estimator{C}^{BB'}_{\ell} =  \frac{1}{2} (\estimator{C}^{+}_{\ell} - \estimator{C}^{-}_{\ell}) \, , \\
     \estimator{C}^{EB'}_{\ell} =  \frac{1}{2} (\estimator{C}^{\times +}_{\ell} + \estimator{C}^{\times -}_{\ell}) \, ,  \\
     \estimator{C}^{BE'}_{\ell} =  \frac{1}{2} (\estimator{C}^{\times +}_{\ell} - \estimator{C}^{\times -}_{\ell}) \, .
\end{gather}

Therefore, we see that the mask correlation function acts as the spectrum of eigenvalues of the mixing matrix. Indeed, in Appendix~\ref{sect:mixing_matrix_eig} we show that the eigenvectors of~$M_{\ell\ell'}$ are the Wigner $d$-matrices which satisfy
\begin{equation}
    \sum_{\ell'} M_{\ell\ell'} d^\ell_{ss'}(\theta) = 2 \pi \xi^{ww'}(\theta) d^\ell_{ss'}(\theta) \;,
\end{equation}
with an associated eigenvalue~$\xi^{ww'}(\theta)$. Most importantly, in Eq.~\eqref{eq: All} we can see that the mixing matrix will be invertible if and only if
\begin{equation} \label{eq: master_condition}
    \xi^{ww'}(\theta) > 0 \quad \forall \; \theta \in [0, \pi] \; ,
\end{equation}
or, in other words, if all the eigenvalues of the mixing matrix are bigger than zero.

Thus, we have shown that when Eq.~\eqref{eq: master_condition} holds, unmixing strategies in either harmonic or real space are actually equivalent. However, for small enough masks there is an effective maximum angular separation $\theta^*$ beyond which the number of pairs is either extremely low or zero. In such cases, the inverse of the correlation function of the mask diverges, preventing us from estimating the full-sky angular power spectra in a straightforward manner. In the next section we will see how different strategies to cope with this fact lead to effectively different pseudo-$C_\ell$ estimators.

Finally, we note that in Eq.~\eqref{eq:master} and thereafter we have assumed sums over an infinite number of modes. However, in practice there is an $\ell_{\rm max}$ up until which the harmonic coefficients of the maps are measured, which is itself tied to the resolution of maps generated from the galaxy catalogue. This $\ell_{\rm max}$ introduces a bias at high-$\ell$ in our estimates due to the down-mixing of power from scales beyond $\ell_{\rm max}$. In what follows, we will assume that our measurements are non-band limited since this problem is common to all estimators and is independent of the footprint of the mask. In Sect.~\ref{Sect: data} we describe how we bypass this problem.

\subsection{\texttt{MASTER} -- The harmonic space solution}\label{Sect: NaMaster}
The \texttt{MASTER} methodology addresses the problem laid out in Sect.~\ref{sect: NaturalSpice} by ingeniously sidestepping it. The \texttt{MASTER} methodology does so by binning the mixing matrix prior to inverting it. This effectively bunches together the problematic eigenvalues, making the binned mixing matrix invertible  \citep{0105302, 1809.09603, improved_Master}. However, because binning is a non-invertible operation, the \texttt{MASTER} methodology cannot be expressed as a linear operator acting on Eq.~\eqref{eq:ev-cl-mm}. Instead, the \texttt{MASTER} estimator is given by
\begin{equation} \label{eq:master_2}
    \estimator{C}^{f\!f'}_q  = \sum_{q'} (\BinnedM{}^{ww'})^{-1}_{qq'} \tilde{C}_{q'} \, ,
\end{equation}
where $\BinnedM{}_{qq'}^{ww'} = B_{\ell q} M_{\ell\ell'}^{ww'} B_{\ell'q'}$ is the binned mixing matrix resulting from the binning scheme of Eq.~\eqref{eq:binning}. Implicit in Eq.~\eqref{eq:master_2} is an assumption that the observed spectra are constant within bins. Thus, if taken at face value, the \texttt{MASTER} estimator is a biased estimator of the binned full-sky angular power spectra. In order to address this bias, the theoretical expectations for the full-sky spectra must be multiplied by the binning kernel 
\begin{equation} \label{eq: namaster_kernel}
    K^{w\!w'}_{q\ell}  = \sum_{q'} (\BinnedM{}^{w\!w'})_{qq'}^{-1} \sum_{\ell'} B_{\ell' q'}M_{\ell\!\ell'}^{ww'} \,,
\end{equation}
before being compared to the \texttt{MASTER} estimates. Moreover, the \texttt{MASTER} algorithm is only capable of producing estimates of the binned power spectra. It is, nonetheless, a remarkably robust and simple methodology, as we will see. 

The \texttt{MASTER} methodology can be directly applied to auto- and cross-correlation spectra of spin-2 fields. In addition to this, \citet{1809.09603} provides a map-level `purification' routine to prevent the mixing of power between the $E$- and $B$-mode components due to binning. The idea behind this algorithm is to undo the impact of the mask directly on the maps, before computing their associated harmonic coefficients. However, doing so requires the mask to be differentiable, having to be convolved with an apodisation beam to smooth its edges. This apodisation procedure effectively changes the footprint being deconvolved. For this reason, we disfavour this approach.   
 
\subsection{\texttt{PolSpice} -- The real space solution} \label{Sect: PolSpice}
The \texttt{PolSpice} algorithm addresses the problem laid out in Sect.~\ref{sect: NaturalSpice} head on by building a pseudo-inverse of the correlation function of the mask. In other words, the \texttt{PolSpice} methodology is effectively equivalent to performing pseudo-inversion of the mixing matrix via singular value decomposition. Thus,  Eq.~\eqref{eq: All} becomes
\begin{equation} \label{eq: pseudo-All}
    A^{\rm apo}_{\ell\ell'} = \frac{2\ell'+1}{2} \int^{\pi}_{0} \frac{h(\theta)}{\xi^{ww'}(\theta)} \, d^{\ell}_{ss'}(\theta) \, d^{\ell'}_{ss'}(\theta) \, \sin\theta \, \diff\theta \, ,
\end{equation}
where $h(\theta) \in [0, 1]$ is the so-called apodisation function that suppresses the inverse of $\xi^{ww'}(\theta)$ when it diverges. Furthermore, $h(\theta)$ must be smooth to avoid ringing when transforming between harmonic and real space. 

Thus, $h(\theta)$ can be understood as effectively limiting the angular range in which the inverse of the correlation function is defined up to a maximum angular separation, $\theta^*$, beyond which it is set to zero. In doing so $h(\theta)$ breaks the orthogonality relationship between the Wigner $d$-matrices that ensures that the estimators for the unmixed angular power spectra defined in Sect. \ref{sect: NaturalSpice} are unbiased.

Thus, \texttt{PolSpice} is also a biased estimator of the full-sky angular power spectra. This bias is given by the kernel 
\begin{equation} \label{eq: Hll}
    H_{\ell\ell'} \equiv \frac{2\ell'+1}{2} \int_0^\pi h(\theta) \, d^\ell_{ss'}(\theta) \, d^{\ell'}_{ss'}(\theta) \, \sin\theta \, \diff\theta \; ,
\end{equation}
which quantifies how apodisation breaks the orthogonality of the Wigner $d$-matrices. Unlike in the case of the \texttt{MASTER} algorithm, where the kernel is applied to the theory expectation to forward model the impact of the bias, \citet{0303414} proposed removing the bias by dividing the \texttt{PolSpice} estimate by 
\begin{equation} \label{eq: hll}
    h_\ell \equiv \sum_{\ell'} H_{\ell \ell'} \; ,
\end{equation}
which measures the response of a flat power spectrum to the apodisation function. Thus, the \texttt{PolSpice} estimates can be directly compared to the theory expectations. 

For auto-correlations of spin-0 fields and cross-correlations of spin-0 and spin-2 fields these biases suppress the power of the angular power spectra at low-$\ell$. However, for auto-correlations of spin-2 fields apodisation also leads to power being leaked between the $E$- and $B$-mode components of the fields, much like masking does \citep{0303414}. We explicitly show this mixing of power taking place in Appendix~\ref{sect: Purification}. As described by \citet{10.1051} and \citet{Crittenden_2002}, it is possible to remove this leakage of power by constructing pure $E$- and $B$-mode correlation functions. To do this, we construct the operator 
\begin{gather} 
 [T(f)](\theta_{x})  =  f(\theta_{x})  \nonumber\\
\qquad {} - \frac{8(2+x)}{(1-x)^2} \int_x^1 \frac{1-x'}{(1+x')^2}  \, f(\theta_{x'})  \, \diff x'  \nonumber\\ 
\qquad {} + \frac{8}{1-x} \int^1_x \frac{1}{(1+x')^2}  \, f(\theta_{x'})  \, \diff x'
\, , \label{eq:T}
\end{gather}
and its inverse
\begin{gather} 
 [U(f)](\theta_{x})  =  f(\theta_{x})  \nonumber\\
\qquad {} - \frac{8(2-x)}{(1+x)^2} \int_{-1}^{x}  \frac{1+x'}{(1-x')^2}  \, f(\theta_{x'})  \, \diff x'   \nonumber\\
\qquad {} + \frac{8}{1+x} \int_{-1}^{x} \frac{1}{(1-x')^2}  \, f(\theta_{x'})  \, \diff x' \, ,
\label{eq:U}
\end{gather}
where $\theta_x = \arccos(x)$ and $f(\theta)$ is a generic angular function. These operators are defined such that $T[d^\ell_{22}](\theta) = d^\ell_{2-2}(\theta)$ and $U[d^\ell_{2-2}](\theta) = d^\ell_{22}(\theta)$.  Thus, the operator $T$ projects a function defined in the basis of $d^\ell_{22}(\theta)$ to the $d^\ell_{2-2}(\theta)$ basis and vice versa for $U$. The derivation of these operators can be found in Appendix~\ref{sect:d_matrix_relations}.  

Therefore, these operators allow us to write pure $E$- and $B$-mode correlation functions by considering the linear combinations
\begin{equation} \label{eq:xi_E_m}
    {\xi}^{EE'}_-  \equiv \frac{1}{2}\left( T[\xi^+]  +  \xi^-  \right) = \sum_{\ell} \frac{2\ell+1}{4\pi} {C}_{\ell}^{EE'} d^{\ell}_{2-2}(\theta) \; ,
\end{equation}
\begin{equation} \label{eq:xi_E_p}
    {\xi}^{EE'}_+  \equiv \frac{1}{2}\left( \xi^+  +  U[\xi^-] \right) = \sum_{\ell} \frac{2\ell+1}{4\pi} {C}_{\ell}^{EE'} d^{\ell}_{22}(\theta) \; ,
\end{equation}
\begin{equation} \label{eq:B_m}
    {\xi}^{BB'}_-  \equiv \frac{1}{2}\left( T[\xi^+] - \xi^-  \right)  = \sum_{\ell} \frac{2\ell+1}{4\pi} {C}_{\ell}^{BB'} d^{\ell}_{2-2}(\theta) \; ,
\end{equation}
and
\begin{equation} \label{eq:B_p}
    {\xi}^{BB'}_+  \equiv \frac{1}{2}\left( \xi^+ - U[\xi^-]  \right)  = \sum_{\ell} \frac{2\ell+1}{4\pi} {C}_{\ell}^{BB'} d^{\ell}_{22}(\theta) \; .
\end{equation}
Before moving forward, it is worth appreciating that, while each pure-mode correlation function can in principle be computed in two different ways -- either by projecting $\xi^{+}(\theta)$ onto $d^\ell_{2-2}(\theta)$ or $\xi^{-}(\theta)$ onto $d^\ell_{22}(\theta)$ -- only one is feasible given our limitations. In Eq.~\eqref{eq:U} we can see that the $[U(f)](\theta)$ operator requires access to arbitrarily large angular separations, which are precisely the angular scales at which $\estimator{\xi}^{\pm}(\theta)$ diverge. Thus, we cannot in practice project $\estimator{\xi}^{-}(\theta)$ on the $d^\ell_{22}(\theta)$ basis. Therefore, we will from now on only consider ${\xi}^{EE'} \equiv {\xi}^{EE'}_-$ and ${\xi}^{BB'} \equiv {\xi}^{BB'}_-$\footnote{We note that, although possible, the computation of ${\xi}^{EE'}_-$ and ${\xi}^{BB'}_-$ is challenging. This is because of the pole in Eq.~\eqref{eq:T} at $x=1$. At these small angular separations $T[f](\theta) \approx f(\theta)$, which requires the cancellation of the divergent second and third terms in the operator. Thus, careful numerical integration is needed.}.

Armed with these pure $E$- and $B$-mode correlation functions, we can write the following pure estimators for the full-sky angular power spectra without leakage given by
\begin{equation} \label{eq:Cl_EE_pure}
    \estimator{C}_\ell^{EE'}  \equiv \int^\pi_0 h(\theta) \frac{\tilde{\xi}^{EE'}(\theta)}{\xi^{ww'}(\theta)}  d^\ell_{2-2}(\theta) \, \sin\theta \, \diff\theta \; ,
\end{equation}
and
\begin{equation} \label{eq:Cl_BB_pure}
    \estimator{C}_\ell^{BB'}  \equiv \int^\pi_0 h(\theta) \frac{\tilde{\xi}^{BB'}(\theta)}{\xi^{ww'}(\theta)}  d^\ell_{2-2}(\theta) \, \sin\theta \, \diff\theta \; .
\end{equation}
Since $E$- and $B$-mode information is already isolated in the purified correlation functions, Eqs.~(\ref{eq:xi_E_m}) and (\ref{eq:B_m}), these new estimators prevent apodisation from mixing power. They nonetheless still need to be divided by their associated kernel given by Eq.~\eqref{eq: hll} to become fully unbiased.

The final question to answer is what specific form the apodisation function $h(\theta)$ should take. Unfortunately, there is no optimal form for $h(\theta)$. Instead, its shape is determined by whether we prioritise less noisy more correlated estimates or vice versa. Broad apodisation functions, such as the Gaussian functions suggested in \citet{0303414}, 
\begin{equation} \label{eq: gaussian_apo}
    h_{\rm gaussian}(\theta, \theta^*, \lambda) = \mathrm{e}^{-\lambda (\theta/\theta^*)^2} \, ,
\end{equation}
heavily down-weight noisy angular separations at the expense of correlating the modes of the final angular power spectra estimates. Similarly, more abrupt apodisation functions, such as a logistic function
\begin{equation} \label{eq: logistic_apo}
    h_{\rm logistic}(\theta, \theta^*, \lambda) = 1/\left[1+\mathrm{e}^{-\lambda(\theta-\theta^*)}\right] \, ,
\end{equation}
introduce less correlation in the final estimates at the expense of more noise. We will explore the interplay between these two factors in more depth in Sect.~\ref{Sect: Results}. We note that both apodisation functions depend on the maximum angular separation included in the analysis, $\theta^*$, and a steepness parameter, $\lambda$.
 
\begin{figure}
    \centering
    \includegraphics[width=0.85\linewidth]{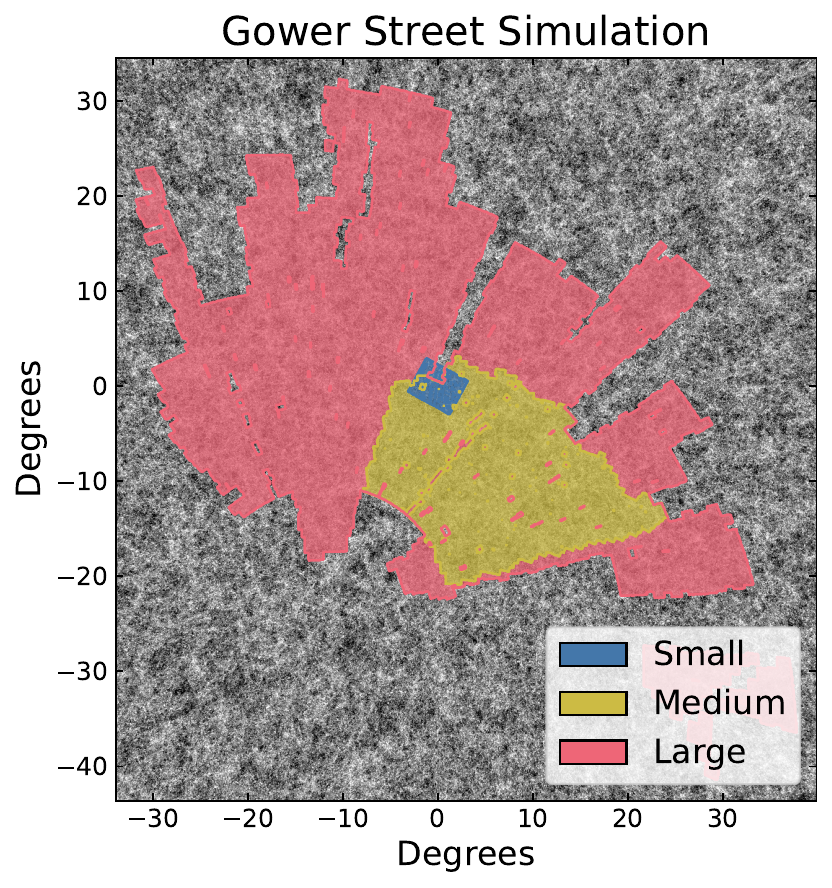}
    \caption{Three masks considered in this work: large ($1615\sqdeg$; red), medium ($378\sqdeg$; yellow), and small ($20\sqdeg$; blue). The background shows one density-field realisation from the GS simulations.}
    \label{fig: masks}
\end{figure}
\begin{figure*}
    \centering
    \includegraphics[width=1\linewidth]{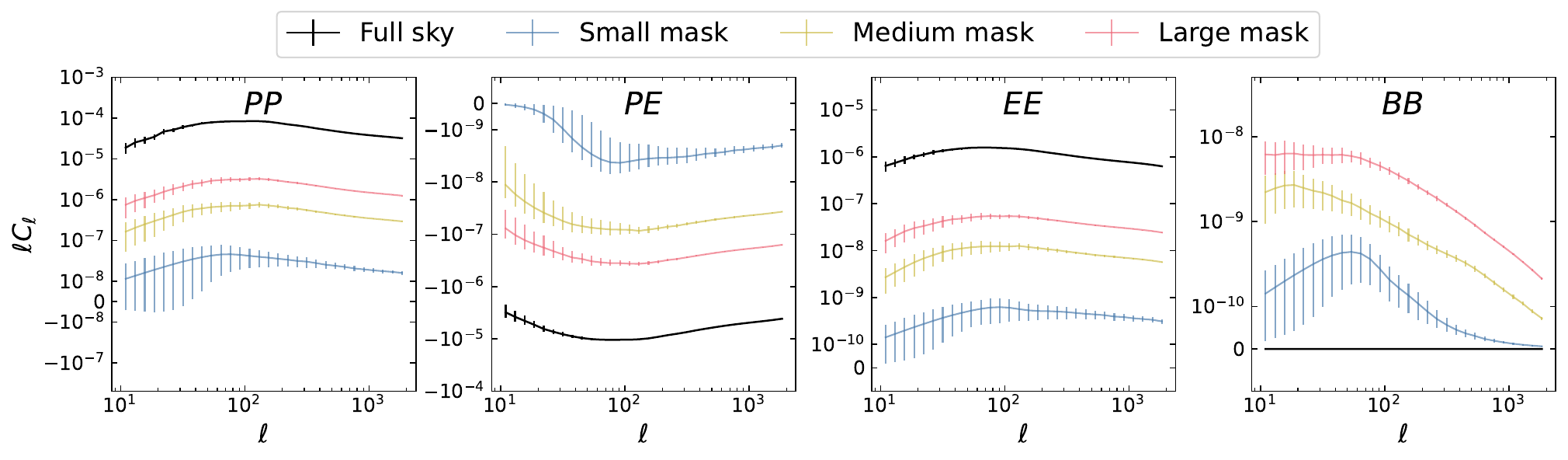}
    \caption{Mean angular power pseudo-spectra measured from the positions and convergence fields of the 200 fixed-cosmology GS simulations used in this work. Left to right, we show the position field auto-correlation ($PP$), the cross-correlation between the position and convergence $E$-mode field ($PE$), the auto-correlation of the convergence $E$-mode field ($EE$), and the auto-correlation of the convergence $B$-mode field ($BB$). In each panel we plot the spectra measured from the full-sky fields as well as when the different masks shown in Fig.~\ref{fig: masks} are applied. We can see that as a larger portion of the sky is masked, the amplitude of the spectra decreases. Moreover, masking the sky results in leakage between the $E$-mode and $B$-mode spectra. Error bars are given by the ensemble covariance of the simulations. }
    \label{fig: mixed_cls}
\end{figure*}
 
\section{Data} \label{Sect: data}
In order to investigate the impact of masking in angular power spectra we make use of the GS suite of $N$-body simulations \citep{GowerSt}. The GS simulations were created using the $N$-body code \texttt{PKDGRAV3} \citep{PKDGRAV} to create realistic lightcones within the simulated box stored in the form of thin shells. In particular we make use of a series of 200 simulations run at the \Euclid Flagship cosmology  \citep{Flagship} that will be released in the upcoming GS 2 suite of simulations. We reweight the simulations' matter shells by an idealised Smail redshift distribution \citep{Smail} given by
\begin{equation}
    n^{\rm smail}(z) \propto \left(\frac{z}{z_{\rm median}}\right)^a \exp\left[-\left(\frac{z}{z_{\rm median}}\right)^b\right] \, , 
\end{equation}
with $z_{\rm median}=1.0$, $a=1.5$, and $b=2.0$. We then split the distribution of galaxies into two equal density tomographic bins assuming a Gaussian error of $\sigma_z = 0.05$ in the assignment of galaxies to each bin. We consider the first bin as the `lens' population distorting the light from the second `source' bin. We then use the \texttt{GLASS} library \citep{2302.01942} to convert the reweighted shells into full-sky convergence and position fields with 3.5 galaxies per square arcminute and 0.26 ellipticity dispersion. 

We then apply three masks, defined so as to allow us to study the impact of sky masking on the measured angular power spectra in different regimes (Fig.~\ref{fig: masks}):
\begin{itemize}
\item{Large ($1615\sqdeg$ footprint), representative of the portion Euclid Data Release 1 (DR1) will cover in the southern hemisphere;}
\item{Medium ($378\sqdeg$ footprint), the typical subsection used for \Euclid data validation;}
\item{Small ($20\sqdeg$ footprint), the average size of individual processing patches in the \Euclid catalogue.}
\end{itemize}

Once the fields have been simulated and masked, we measure the auto- and cross-correlation angular power spectra between the position and convergence fields using the library \texttt{Heracles} \citep{Heracles}. We also measure the angular power spectra of the fields in the absence of any masking. In order to do so we impose a pixelisation on the maps with $N_{\rm side}=2048$ resolution. This resolution allows us to measure the angular power spectra of the previously described masks up to $\ell_{\rm max} = 6000$. To properly account for the up- and down-mixing of modes, we compute the pseudo-$C_\ell$ of the fields up to $\ell_{\rm max} = 4000$ and then truncate the final $\estimator{C}_\ell$ estimate below $\ell_{\rm max} = 2000$. In Fig.~\ref{fig: mixed_cls} we show the mean angular power spectra for the auto- and cross-correlations of the simulated position and convergence fields. In this plot we can clearly observe the effect of masking on the measured angular power spectra. In each panel we can see that masking decreases the amplitude of the masked angular power spectra by roughly the fraction of the sky being covered. More interestingly, we can observe the expected leakage of power between the $E$- and $B$-mode spectra. Thus, even if no signal is present in the full-sky $B$-mode spectra, the masked spectrum presents a strong signal. The error bars present in this plot are given by the ensemble covariance of the simulations.   

Finally, note that all the angular power spectra in this work have been estimated from synthetic maps for simplicity. However, in Appendix~\ref{sect: shot noise} we show that the implementation of this methodology inside \texttt{Heracles} can be seamlessly applied to catalogue analyses.
 
\section{Results} \label{Sect: Results}
\begin{figure}
    \centering
    \includegraphics[width=\linewidth]{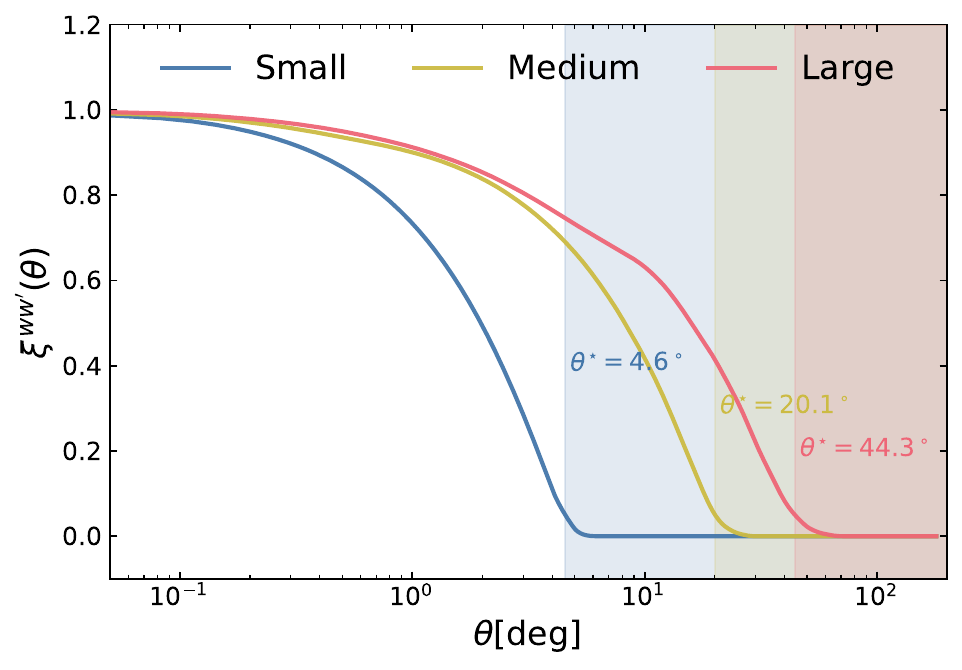}
    \caption{Two-point correlation function of the small (blue), medium (yellow), and large masks (red). The shaded regions show the angular scales suppressed during the inversion of the correlation function of the different masks. The values of $\theta^*$ show the cut-off scale of the suppression for each mask. As the size of the footprint decreases, the values of its correlation function fall to zero at smaller angular separation.}
    \label{fig: regularisation}%
\end{figure}
 
\begin{figure*}
    \centering
    \includegraphics[width=\linewidth]{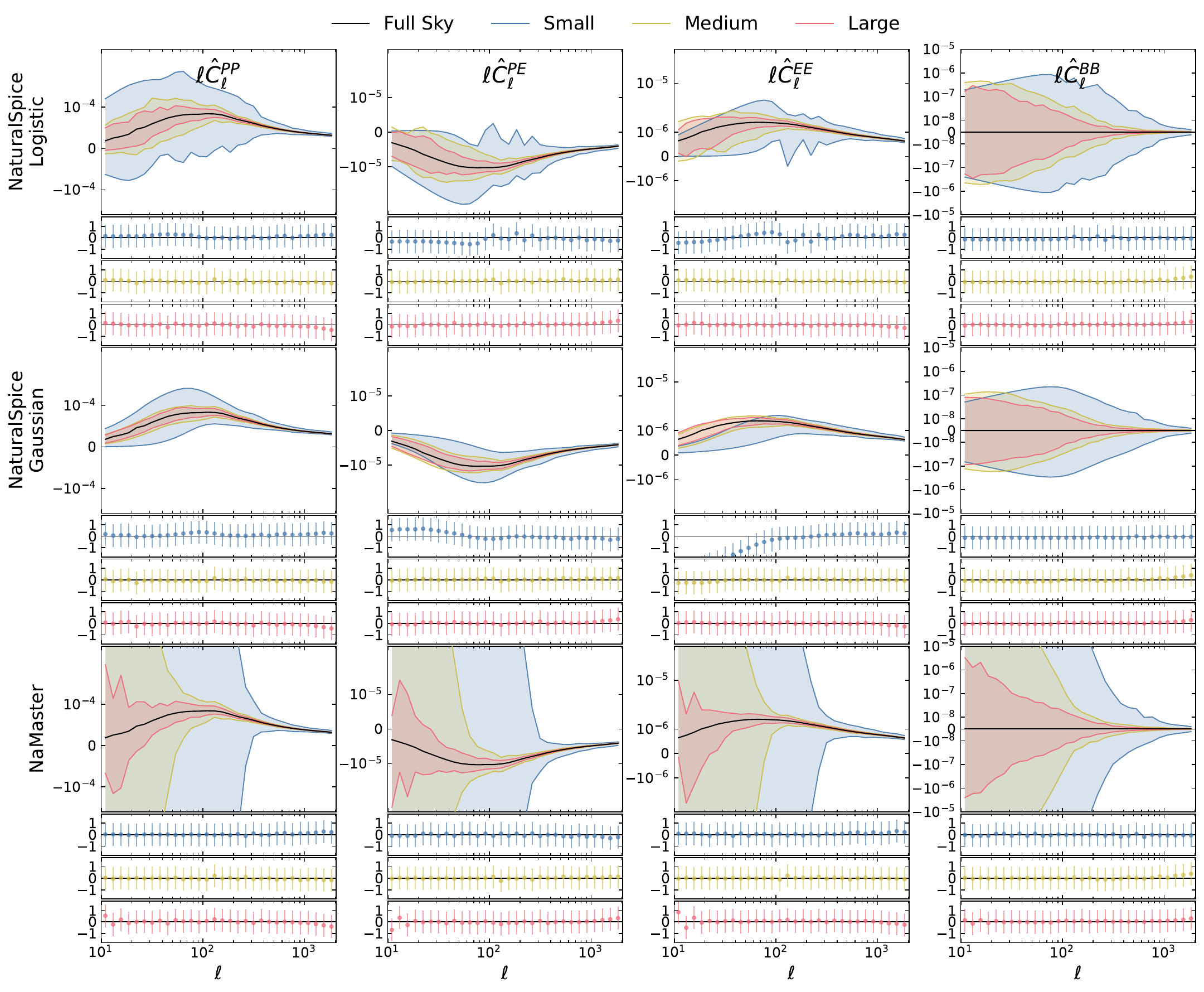}
    \caption{Deconvolved full-sky angular power spectrum for the auto- and cross-correlations of different fields given the small (blue), medium (yellow), and large (red) mask considered in this work. First column shows the galaxy clustering auto-correlation spectra. Second column shows the cross-correlation of the galaxy clustering with the $E$-mode of the shear field. The third column shows the auto-correlation spectra of the $E$-mode of the shear field. The fourth column shows the auto-correlation spectra of the $B$-mode of the shear field. Each row shows one of the different deconvolution algorithms considered in this work: \texttt{NaturalSpice}, using either a logistic or Gaussian apodisation function, and \texttt{NaMaster}. Moreover, each panel shows the theoretical full-sky expectation for the spectra in black. In addition, we include subpanels where the corresponding residuals of each deconvolved spectrum are shown.}
    \label{fig:cls}
\end{figure*}

\subsection{Set-up and regularisation choices} \label{Sect: setup}
We will now use the \texttt{MASTER} and \texttt{PolSpice} algorithms to estimate the full-sky angular power spectra from each of the footprints described in Sect.~\ref{Sect: data}. In the case of the \texttt{MASTER} methodology, we use the public \texttt{Python} implementation of the algorithm, \texttt{NaMaster} \citep{1809.09603}. For \texttt{PolSpice}, we develop our own \texttt{Python} implementation of the original \texttt{Fortran} algorithm, which for the sake of distinction we will call \texttt{NaturalSpice}. \texttt{NaturalSpice} builds on the original \texttt{PolSpice} by allowing users complete control over the apodisation scheme applied to the data. Implementation wise, \texttt{NaturalSpice} uses matrix operations to perform the transformations between real and harmonic space. This form of parallelisation, alongside using \texttt{C} libraries to perform the most expensive parts of the computation, make \texttt{NaturalSpice} the fastest and easiest to use implementation of the \texttt{PolSpice} algorithm. Moreover, we validated that \texttt{NaturalSpice} obtains the same numerical values as the original \texttt{Fortran} code up to a thousandth of a per cent. 

Leveraging the freedom offered by \texttt{NaturalSpice} in terms of apodisation scheme, we will consider two different set ups, one where a Gaussian apodisation was used to regularise the correlation function of the different masks (Eq.~\ref{eq: gaussian_apo}), and another where a logistic function was used instead (Eq.~\ref{eq: logistic_apo}). As described in Sect.~\ref{sect: unmixing}, all the unmixing methods compared in this work apply a form of regularisation to the eigenvalues of the mixing matrix to make the impact of masking invertible. On the one hand, \texttt{NaMaster} bins the mixing matrix before performing the inversion. This effectively averages small and large eigenvalues. To this effect we employ a binning scheme of 30 bins equally spaced in $\log_{10}\ell$ over the range $10 \le \ell \le 2000$ which we apply to the estimate of all algorithms. We also bin the expected full-sky signal to prevent having to deconvolve the \texttt{NaMaster} estimate with its binning kernel (Eq.~\ref{eq: namaster_kernel}). On the other hand, \texttt{PolSpice} performs a pseudo-inverse beyond a certain angular separation. One of the main advantages of this methodology is that this angular scale can easily be motivated by inspecting the correlation function of the mask. In Fig.~\ref{fig: regularisation} we show the correlation function associated with three masks considered in this work. For each mask we can observe that their values vanish for large enough angular scales. Heuristically, the objective is to keep as much of the angular range as possible while introducing the least amount of noise in the reconstruction. In this work we choose $\theta^*/{\rm deg} \in \{4.6, 20.1, 44.3\}$ for the small, medium, and large masks, respectively. These values correspond to the angles at which the correlation functions drop below 5 per cent of their maximum value. Furthermore, both logistic and Gaussian apodisation functions require the user to set the level of steepness of the correction to avoid ringing effects. Following \citet{0303414}, we set the steepness of the Gaussian apodisation function to $\lambda=24\ln 2$. In the case of the logistic apodisation function, we chose the conservative value $\lambda=10$ to safely avoid ringing effects for the considered footprints. It is important to stress that in all three cases we never apodise mask maps, $w(\nv)$, directly -- preserving the footprints.  

\subsection{Unmixed angular power spectra} \label{Sect: unmixed_cls}
Having described the different regularisation schemes employed by each unmixing algorithm, we now study the resulting unmixed angular power spectra. Starting from the set of full-sky maps generated from the 200 realisations in the set of GS simulations, we apply each of the three masks shown in Fig.~\ref{fig: masks} to derive partial-sky maps. We then measure the angular power spectra of both the masks and the partial-sky fields (see Fig.~\ref{fig: mixed_cls}). The resulting pseudo-$C_\ell$ for each realisation are then fed to the different unmixing algorithms to obtain an ensemble of estimates of the full-sky angular power spectra from each mask. This allows us to derive a mean and a covariance for each algorithm's estimate of the full-sky angular power spectra.
 
In Fig.~\ref{fig:cls} we show the deconvolved full-sky angular power spectra reconstructed by the three considered methods. Each column corresponds to different auto- and cross-correlations spectra while different rows show the results for the three deconvolution algorithms considered in this work. Inside each panel we superpose the deconvolved spectra from each of the considered masks as well as the theoretical expectation for the full-sky spectra in black. Moreover we include a subpanel where the residuals of each method relative to its corresponding error bars are shown. For visual clarity, we choose to show the three signal-dominated spectra of our analysis; $C_\ell^{P\!P}$, $C_\ell^{P\!E}$, and $C_\ell^{E\!E}$, alongside one null spectrum, $C_\ell^{B\!B}$. This allows us to validate whether the deconvolution algorithms can reconstruct the expected cosmological signals without introducing systematic effects in the null spectra. We note that $C^{EE'}_\ell$ and $C^{BB'}_\ell$ estimates in this figure were obtained using the purified estimators, Eqs.~(\ref{eq:Cl_EE_pure}) and (\ref{eq:Cl_BB_pure}). In Appendix~\ref{sect: Purification} we discuss the impact of purification in our estimates.

\begin{figure}
    \centering
    \includegraphics[width=1\linewidth]{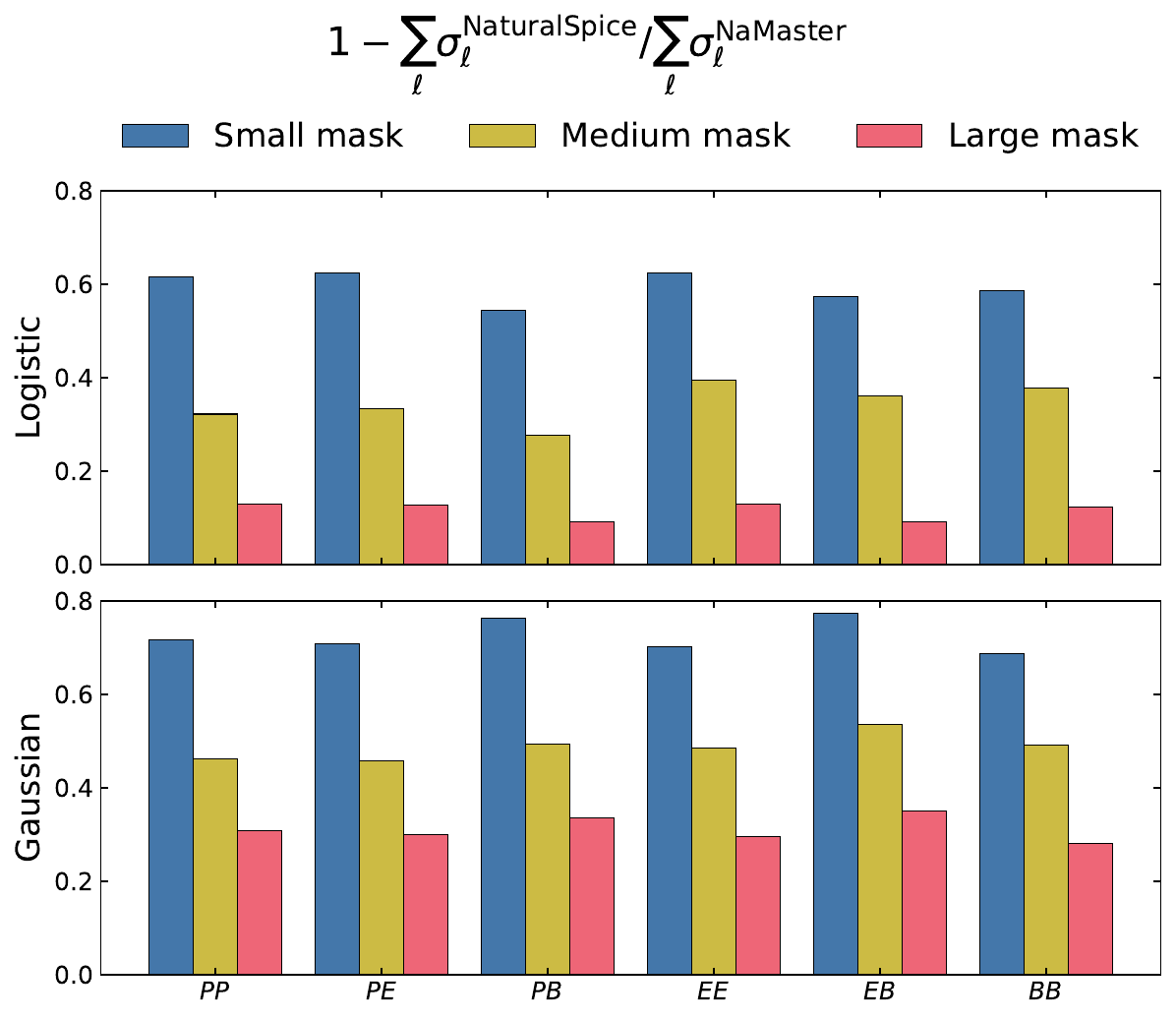} 
    \caption{Relative difference in the variance of $\estimator{C}_\ell$ averaged over the $\ell$ range between \texttt{NaMaster} and \texttt{NaturalSpice}. Top panel shows the results when a logistic function was used to apodise the correlation function of the mask while bottom panel shows the results when a Gaussian apodisation function was used. Different columns refer to different auto- and cross-spectra. Our results show that \texttt{NaturalSpice} yields significantly less noisy angular power spectra estimates, particularly when using a Gaussian apodisation function.  }%
    \label{fig:sigma_ratio}%
\end{figure}

Our results show that even for very pathological masks, such as the small $20\sqdeg$ rectangle, all methods return unbiased estimates for the deconvolved angular power spectra. However, at scales $\ell<100$, we see that for the smallest footprint \texttt{NaturalSpice} apodised with a Gaussian function returns biased estimates of the $\estimator{C}^{EE}_\ell$ in the form of power being suppressed as $\ell$ decreases. This is caused by the tapering of the Gaussian apodisation function being broader than the correlation function of the mask. Thus, the correction laid down in Eq.~\eqref{eq: hll} cannot effectively compensate for the effect of restricting the angular range to such a narrow domain. Most importantly, we observe that more abrupt apodisation functions, such as a logistic function, remain unbiased even in the presence of extremely small masks while avoiding ringing effects. We note that below $\ell=100$ \texttt{NaturalSpice} apodised using a logistic function shows coherent oscillations which are nonetheless compatible with the expected signal. These oscillations are the combined effect of the low numbers of pair counts available to constrain the power spectra at those scales and of Eq.~\eqref{eq: hll}, which correlates the estimates at low-$\ell$. These oscillations are not observable in the \texttt{NaMaster} estimates due to the low S/N at these scales.
 
\subsection{Variance} \label{Sect: variance}
We observe that the variance of different $\estimator{C}_\ell$ estimates depends on what scales we look at. For scales well within the mask, all three deconvolution algorithms generally yield the same amount of variance in the $C_\ell$ estimates regardless of the size of the mask. However, the story radically changes for scales poorly constrained by the mask. At low-$\ell$ \texttt{NaturalSpice} yields orders of magnitude smaller variance than \texttt{NaMaster}. This is true for both signal and null angular power spectra. To better showcase this difference, in Fig. \ref{fig:sigma_ratio} we show the relative difference between the mean variance of the \texttt{NaturalSpice} and \texttt{NaMaster} $\estimator{C}_\ell$ estimates
\begin{equation}
    \ols{\delta\sigma} = 1 - \frac{\sum_\ell \sigma_\ell^{\rm NaturalSpice}}{\sum_\ell \sigma_\ell^{\rm NaMaster}}\; .
\end{equation}
In this metric, numbers closer to one represent \texttt{NaturalSpice} estimates showing less variance. Thus, we can see that, when using a logistic apodisation function, \texttt{NaturalSpice} yields on average 60 per cent less variance than \texttt{NaMaster} in its estimates of angular power spectra when considering a small ($\sim 20 \sqdeg$) mask. As the size of the mask increases the difference between the methods decreases. Hence, we see that for medium ($\sim 400 \sqdeg$) and large ($\sim 1600 \sqdeg$) masks the improvement reduces to approximately 30 and 15 per cent, respectively. Using a Gaussian function instead further reduces the variance of the \texttt{NaturalSpice} estimates. In this case, we observe an average improvement of about 70, 50, and 30 per cent with respect to \texttt{NaMaster}. This is due to the Gaussian apodisation function heavily down-weighting noisy, large angular separations in the correlation function of the mask. As we will see this comes at the cost of more correlated estimates. It is important to stress that in both cases, the improvements are driven by scales $\ell<100$. While these scales are relevant for many science cases, they do not drive the constraints of traditional cosmological parameters.

\begin{table}
\centering
\caption{Mean absolute correlation, $\gamma$, of the ensembles of deconvolved spectra generated by each unmixing algorithm considered in this work (rows) for each of the considered masks (columns). Here $\varrho_{ij}$ are the correlation coefficients of the covariance matrix of the $n$ estimated band powers (Eq.~\ref{eq: gamma}). Smaller values represent more diagonal covariances. We observe that as the size of the mask decreases the degree of correlation $\gamma$ increases in a similar way for all the methods.}\begin{tabular}{p{1.8cm}ccc}
\hline 
\hline
\multicolumn{4}{c}{$\frac{1}{n(n-1)} \sum_{i\neq j}^n |{\rm \varrho}_{ij}|$} \\
\hline
Method & Small mask & Medium mask & Large mask \\
\hline
NaturalSpice Logistic & 0.161 & 0.098 & 0.081 \\
NaturalSpice Gaussian & 0.183 & 0.100 & 0.087 \\
NaMaster & 0.194 & 0.115 & 0.081 \\
\hline
\end{tabular}
\label{tab:diag_score_diff}
\end{table}
 
\begin{figure*}
    \centering
    \includegraphics[width=\linewidth]{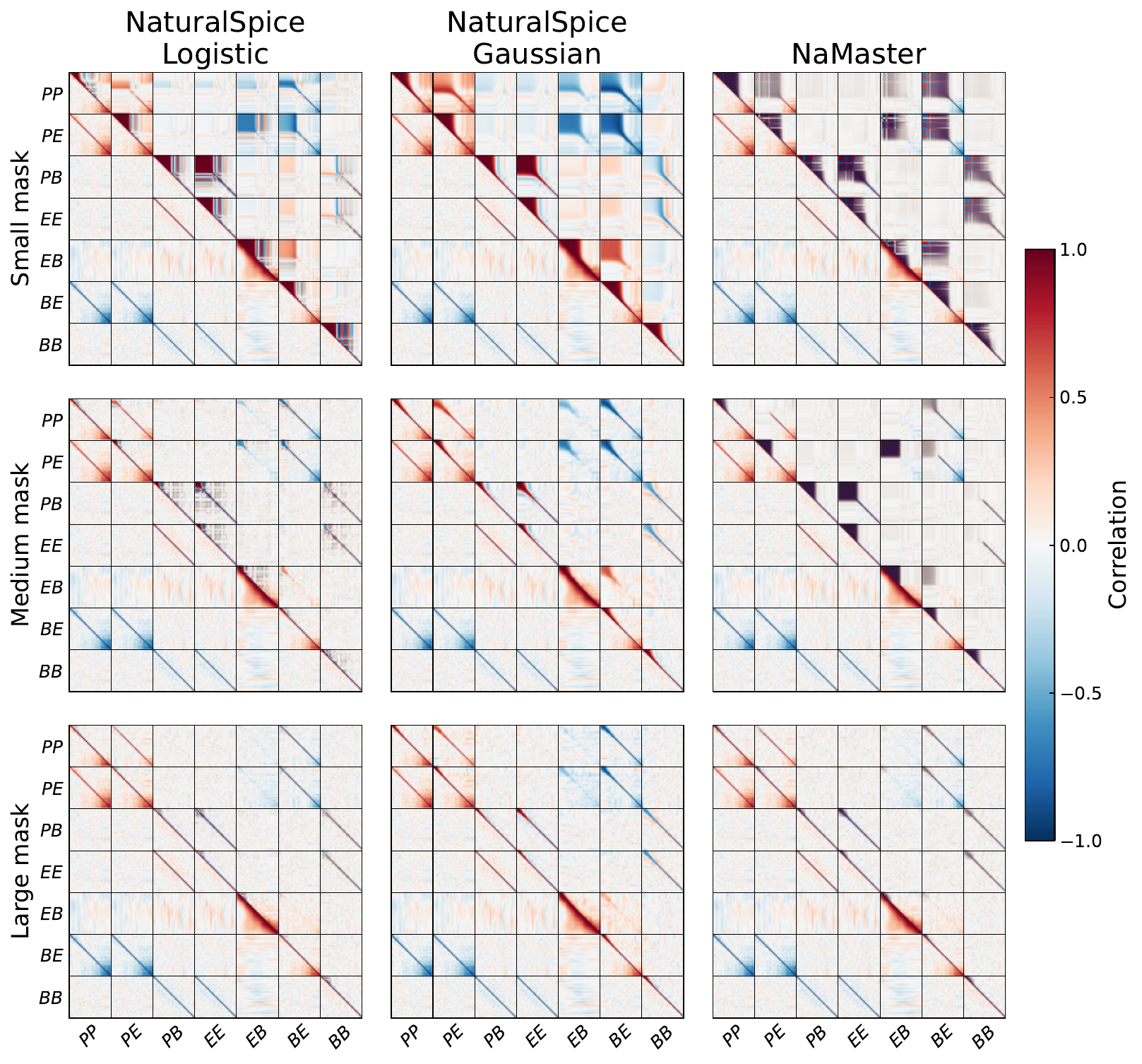}
    \caption{Correlation matrices for the ensemble of deconvolved angular power spectra generated by each of the algorithms considered in this work (columns) when different masks are applied (rows). In each panel the upper triangle shows the correlation matrix of the deconvolved angular power spectra while the lower triangle shows the correlation matrix of the reference ensemble of full-sky spectra for direct visual comparison. We generally observe that, as the mask size decreases the degree of correlation in the reconstructed spectra increases. }
    \label{fig:corr_comp}
\end{figure*}

\begin{figure}
    \centering
    \includegraphics[width=1\linewidth]{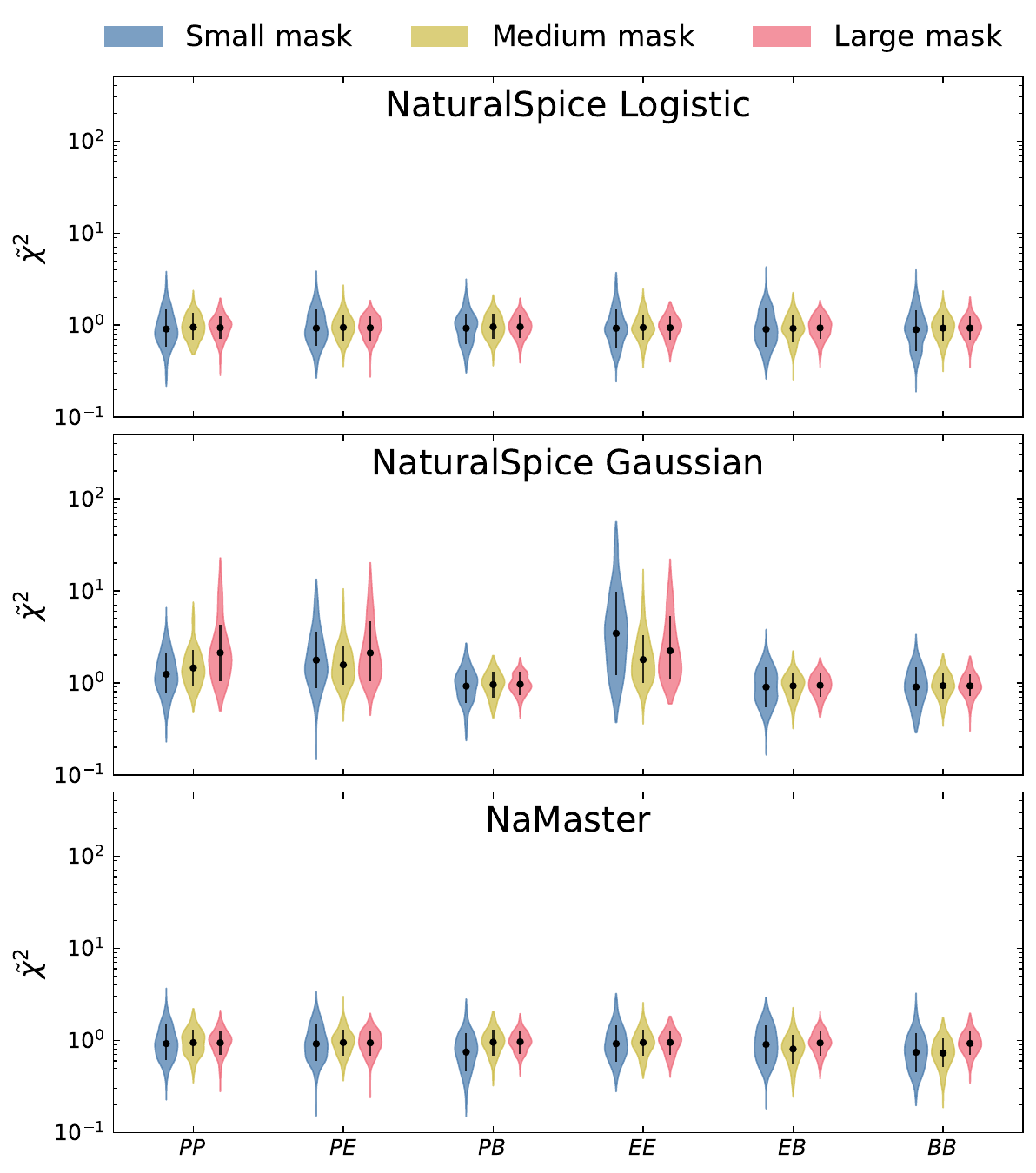}
    \caption{Reduced goodness of fit of each unmixing algorithm computed with respect to a full-sky angular power spectrum, using the covariance of the deconvolved angular power spectra. Different rows show different masks while different columns show different auto- and cross-correlations. We observe that \texttt{NaMaster} and \texttt{NaturalSpice}, apodised with a logistic function, are the only two algorithms consistently returning reduced goodness of fit of order one regardless of the mask applied to the data or the correlation considered.}
    \label{fig:chi2}
\end{figure}
 
\subsection{Correlation} \label{Sect: correlation}
We now turn to the correlation structure of the reconstructed $\estimator{C}_\ell$. In Fig.~\ref{fig:corr_comp} we show the correlation matrix for the estimates of each algorithm (columns) from each mask (rows). To allow a direct comparison, the lower triangle of each correlation shows the same correlation matrix of the full-sky angular power spectra. We see that as the size of the mask decreases, the reconstructed angular power spectra become more correlated. This is particularly noticeable in the low-$\ell$ sections of the individual blocks. In order to quantify this statement and compare the different methods, we compute the absolute mean correlation of each matrix defined as
\begin{equation} \label{eq: gamma}
    \gamma = \frac{1}{n(n-1)} \sum_{i\neq j}^n |\varrho_{ij}| \,,
\end{equation}
where $\varrho_{ij}$ denotes the correlation coefficients of the associated covariance matrices. Thus, if $\gamma$ is zero, the matrix is completely diagonal and if it is one it is completely dense. The computed values of $\gamma$ are shown in Table~\ref{tab:diag_score_diff}. We can see that for all deconvolution methods the level of correlation increases as the mask size decreases. For the largest mask all algorithms report similar levels of correlation in its estimates. However, as the mask decreases \texttt{NaturalSpice} consistently yields lower levels of correlation than \texttt{NaMaster}. Of the two apodisation schemes, the logistic function yields less correlated estimates. In the most extreme scenario, we find that \texttt{NaturalSpice} using a logistic apodisation function yields approximately 20 per cent less correlation than \texttt{NaMaster} for the smallest mask.

Apodisation correlates the angular power spectra estimates through the correction described in Eq.~\eqref{eq: hll}. Broader apodisation functions lead to more severe corrections, which in turn translate into a higher degree of correlation between the reconstructed modes. Thus, less aggressive apodisation functions that preserve the original shape of the mask correlation function, such as a logistic function, yield a lesser degree of correlation. Thus, the choice of apodisation function boils down to a traditional exchange of variance versus correlation. In other words, the mass in the covariance matrix is preserved, it is just distributed differently.  
 
\subsection{Goodness of fit} \label{Sect: chi2}
In order to better understand the interplay between the bias, variance, and correlation of the different algorithms, we compute the reduced goodness-of-fit distributions for each set of full-sky angular power spectra estimates given the measured full-sky $C_\ell$ and the covariance of the estimate,
\begin{equation}
    \tilde{\chi}^2 = \frac{(\mathbfit{d} - \estimator{\mathbfit{d}})^{\tp} \, \estimator{\mathbfss{C}}^{-1} \, (\mathbfit{d} - \estimator{\mathbfit{d}})}{\mathrm{dof}}  \, ,
\end{equation}
where $\mathbfit{d}$ and $\estimator{\mathbfit{d}}$ are the data vectors whose entries are the full-sky and estimated angular power spectra, $C_\ell$ and $\estimator{C}_\ell$, respectively, $\estimator{\mathbfss{C}}$ is the covariance matrix of the estimates, and dof is the number of degrees of freedom. From Fig.~\ref{fig:corr_comp} it is clear that the covariance matrices of the angular power estimates become non-invertible below a certain scale, due to the high degree of correlation. To address this problem we exclude scales below $\ell_{\rm min} \in \{15, 30, 100\}$ for the large, medium, and small masks, respectively.
The mean reduced $\tilde{\chi}^2$ of each ensemble of $\estimator{C}_\ell$ with its associated 68 per cent quantiles can be found in Fig.~\ref{fig:chi2}. Looking at the computed values, we can observe that \texttt{NaturalSpice}, when using a logistic apodisation function,  and \texttt{NaMaster} consistently return reduced $\chi^2$ of order one for all the auto- and cross-correlations of the data. We observe this for all the masks considered in this work, as well as for signal ($PP$, $PE$, $EE$) and null ($PB$, $EB$, $BB$) auto- and cross-correlations. When using a Gaussian apodisation function instead, \texttt{NaturalSpice} yields significantly higher $\tilde{\chi}^2$ values due to a combination of the biases noted in Fig.~\ref{fig:cls} and the higher degree of correlation, making the inverse covariance noisier.

\subsection{Computational performance} \label{Sect: speed}
Concerning speed, we observe that \texttt{NaturalSpice} is by far the fastest algorithm. For a precomputed set of correlation functions, \texttt{NaturalSpice} deconvolves the whole set of pseudo-$C_\ell$ in an average time of 7.6 seconds. The second fastest algorithm is \texttt{NaMaster}. For a pre-computed mixing matrix, \texttt{NaMaster} takes 10.34 seconds on average. Finally, \texttt{PolSpice} is by far the slowest algorithm, requiring 98.43 seconds on average. All of these benchmarks were produced using one single CPU node on the Perlmutter supercomputer at NERSC.
 
\section{Conclusions} \label{Sect: Conclusions}
 
\paragraph{\textbf{What is \texttt{NaturalSpice}?}} 
\texttt{NaturalSpice} is a new \texttt{Python} implementation of the \texttt{PolSpice} algorithm based on reformulating pseudo-$C_\ell$ estimation as a pseudo-inversion problem. In this picture, the apparent mixing of harmonic modes induced by the mask is understood as the consequence of working in a basis where the masking operator is not diagonal. We show that the basis that diagonalises the mask is real space, where the correlation function of the mask gives the eigenvalues of the mixing matrix. As a result, harmonic- and real-space unmixing strategies are equivalent in the absence of a band limit. In practice, however, working in real space makes it possible to regularise the non-invertible part of the problem in a controlled way. The amount of regularisation required by a given footprint can be read directly from its correlation function, allowing users to choose it in a transparent way. Moreover, unlike the \texttt{MASTER} methodology, which can only produce estimates of the binned power spectra, \texttt{NaturalSpice} is not restricted to binned estimates. \texttt{NaturalSpice} is also fast, writing the transformations between harmonic and real space as matrix operations and outsourcing the expensive convolution integrals to \texttt{C} libraries.

\paragraph{\textbf{When should you use \texttt{NaturalSpice}?}} 
We benchmarked \texttt{NaturalSpice} against the current state-of-the-art library, \texttt{NaMaster}, using 200 fixed-cosmology realisations of the GS simulations. We measured full-sky and masked power spectra for galaxy positions and weak lensing convergence fields, and compared the reconstructed full-sky spectra obtained from three representative masks: a large $1615\sqdeg$ mask, a medium $378\sqdeg$ mask, and a small $20\sqdeg$ patch. We evaluated the methods in terms of the bias, variance, and correlation of their estimates.
 
For the large mask, the performance of the two methods is broadly comparable, with \texttt{NaturalSpice} yielding approximately 15 per cent lower variance and the same level of correlation as \texttt{NaMaster}. As the footprint decreases, the advantage of \texttt{NaturalSpice} grows: for the medium mask it yields approximately 30 per cent lower variance, and for the small mask approximately 60 per cent lower variance, alongside approximately 20 per cent less correlated estimates. These gains are concentrated at $\ell < 100$, the scales most impacted by masking. \texttt{NaturalSpice} achieves this while remaining unbiased.

The choice of apodisation function controls a trade-off between variance and correlation. A Gaussian apodisation function reduces the variance further, but at the cost of more correlated estimates and a suppression of power at the lowest multipoles for the smallest footprint. We therefore recommend the use of a logistic apodisation function, which in our tests remains unbiased for all footprints, yields the least correlated estimates of the methods considered, and consistently returns reduced $\chi^2$ values of order one.
 
The additional constraining power brought about by \texttt{NaturalSpice} will allow cosmological analyses to improve their constraints on parameters determined by large-scale information such as the primordial power spectrum amplitude and tilt, $A_{\rm s}$ and $n_{\rm s}$, respectively, and to better control systematic effects on these scales. However, we do not foresee meaningful gains on parameters such as $\Omega_{\rm m}$ and $\sigma_{\rm 8}$, whose constraints are dominated by small scales. 

We expect \texttt{NaturalSpice} to be particularly valuable for pseudo-$C_\ell$ analyses of small or challenging survey footprints, such as early data releases of Stage-IV surveys or the study of narrow, deep fields, as well as a tool to check for large-scale systematic effects. For larger footprints, \texttt{NaturalSpice} provides a competitive alternative to current algorithms with a transparent regularisation scheme. 
 
\paragraph{\textbf{Future work}} 
When considering joint galaxy clustering and cosmic shear analyses with correlated tomographic bins, the clustering of galaxies contributes a four-point term to the cosmic shear power spectrum, known as the source-lens clustering (SLC) term \citep{2025A&A...693A.210L}. \citet{Heracles} show that SLC manifests as an effective high-$\ell$ bias in both map- and catalogue-based analyses. Since the \texttt{NaturalSpice} methodology divides the cosmic signal by the correlation function of the visibility field, which carries information about the clustering of sources, future work could study whether spectra deconvolved using \texttt{NaturalSpice} suffer from the same SLC bias.

It would also be interesting to compare the performance of \texttt{NaturalSpice} to statistical pseudo-$C_\ell$ estimators, that infer the full-sky angular power spectrum from a given set of partial-sky observations. For example, the quadratic maximum likelihood estimator optimises for the full-sky angular power spectrum that best describes the variance of map's pixels \citep{Tagmark_QML, Maraio_QML}. Similarly, hierarchical approaches have been proposed that promote the optimisation process to a full Bayesian inference problem on the harmonic coefficients of the map \citep{Alsing_Almanac, Sellentin_Almanac}. Both of these statistical approaches yield tighter constraints on the full-sky angular power spectrum at large scales, the ones most impacted by the mask, than traditional unmixing approaches. However, this is achieved at the expense of speed since the dimensionality of the statistical problem grows with the square of the resolution of the maps.
 
Finally, given its speed and its lower variance at large scales, future work could explore implementing \texttt{NaturalSpice} inside simulation-based inference frameworks, where both properties are paramount.

\paragraph{\textbf{Software availability}}
A \texttt{Python} implementation of \texttt{NaturalSpice} is available in the \texttt{Heracles} library (\url{https://github.com/heracles-ec/heracles}), where a tutorial can be found.

\begin{acknowledgements}
The authors would like to thank and remember the late Karim Benabed for his invaluable comments that played a vital role in the conceptualisation of this work. We would also like to thank Vincenzo Cardone for the attentive and insightful review of the manuscript.
JRZ, and BJ acknowledge support by the ERC-selected UKRI Frontier Research Grant EP/Y03015X/1 and by STFC grant ST/W001721/1.
The Gower Street simulations were generated using the STFC DiRAC High Performance Computing Facility (\url{www.dirac.ac.uk}).
This research used resources of the National Energy Research Scientific Computing Center, a DOE
Office of Science User Facility supported by the Office of Science of the U.S. Department of Energy under Contract No. DE-AC02-05CH11231 using NERSC award HEP-ERCAP0037257.
For the purpose of Open Access, the authors have applied a CC BY public copyright licence to any Author Accepted Manuscript version arising from this submission.
We made extensive use of the {\tt numpy} \citep{van2011numpy}, {\tt scipy} \citep{scipy}, {\tt astropy} \citep{astropy_2013, astropy_2018}, {\tt healpy} \citep{Zonca2019}, {\tt GetDist} \citep{getdist}, and {\tt matplotlib} \citep{Hunter_2007} python packages. 
\par \AckEC \AckDRone

\end{acknowledgements}

\bibliography{main}

\begin{appendix}
 
\section{Mixing matrix eigenvalues} \label{sect:mixing_matrix_eig}
The formal eigenvectors of~$M_{\ell\ell'}$ are angular power spectra~$V_\ell$ that satisfy
\begin{equation}
    \sum_{\ell'} M_{\ell\ell'} \vec{V}_{\ell'} = \lambda \vec{V}_\ell \;,
\end{equation}
where~$\lambda$ is the eigenvalue of~$\vec{V}$.  In real space, the same equation for the angular correlation functions is much simpler,
\begin{equation}
    \xi^{ww'}(\theta) \, V(\theta) = \lambda \, V(\theta) \;.
\end{equation}
The only non-trivial functions that satisfy this equation are the Dirac delta functions,
\begin{equation}
    \xi^{ww'}(\theta) \, \delta^{\rm D}(\theta - \theta_k)= \xi^{ww'}(\theta_k) \, \delta^{\rm D}(\theta - \theta_k) \;,
\end{equation}
where the angles~$\theta_k$ are given by some indexing scheme. Using the representation of the delta functions in terms of the Wigner $d$-matrices for the spins of the fields,
\begin{equation}
\sum_\ell \frac{2\ell+1}{2} d_{ss'}^\ell(\theta) d^\ell_{ss'}(\theta_k) = \delta^{\rm D}(\theta-\theta_k) \, ,
\end{equation}
we can simply read off the eigenvector~$V_\ell^{(k)}$ associated with the angle~$\theta_k$,
\begin{equation}
\label{eq:eigvec}
    V_\ell^{(k)} = 2\pi \, d^\ell_{ss'}(\theta_k) \;,
\end{equation}
and its eigenvalue is~$\xi^{ww'}(\theta_k)$.
Correctness of this result is readily checked by the direct computation
\begin{gather}
    \sum_{\ell_1} M_{\ell\ell_1} \, 2\pi \, d^{\ell_1}_{ss'}(\theta_k) \nonumber \\
    = (-1)^{s-s'} \sum_{\ell_1 \ell_2} \frac{(2\ell_1+1) \, (2\ell_2+1)}{2} \,
        \threej{\ell_1,s_1}{\ell_2,s_2}{\ell,-s} \nonumber \\
    \quad \times \threej{\ell_1,s_1'}{\ell_2,s_2'}{\ell,-s'} \, C_{\ell_2}^{ww'} \, d^{\ell_1}_{ss'}(\theta_k) \nonumber \\
    = \sum_{\ell_2} \frac{2\ell_2 + 1}{2} \, C_{\ell_2}^{ww'} \,d^{\ell_2}_{ss'}(\theta_k) \, d^{\ell}_{ss'}(\theta_k) \nonumber \\
    = \xi^{ww'}(\theta_k) \, 2\pi \, d^{\ell}_{ss'}(\theta_k) \;.
\end{gather}
Here the first equality uses the definition~\eqref{eq:mm} of the mixing matrix, the second equality uses the addition theorem for Legendre polynomials, and the third equality uses the relation between angular power spectrum and angular correlation function. 
 
We have shown that diagonalising the mixing matrix and computing the correlation function of the mask are equivalent operations. However, the previous derivation relies on the mixing matrix not being band-limited, meaning that $\ell_{\rm max}=\infty$, whereas in practice the mixing matrix always has a finite size. Thus, computing the spherical harmonics explicitly is more accurate than doing so by diagonalising the mixing matrix.

\section{Wigner $d$ matrix relations} \label{sect:d_matrix_relations}

Our goal is to relate $d_{2,2}^\ell$ and $d_{2,-2}^\ell$; we follow Eqs.~(87)--(90) of \citet{0303414}. Using the recursive properties of the Wigner matrices \citep[][sections 4.8 and 4.9]{Varshalovich}, and writing $\theta_{x} \equiv \arccos(x)$, we obtain
\begin{gather}
d_{2,-2}^\ell(\theta_{x}) =  d_{2,2}^\ell(\theta_{x}) \nonumber\\
\qquad {} - \frac{8(2+x)}{(1-x)^2} \int_{x}^{1} \frac{1-x'}{(1+x')^2} d_{2,2}^\ell(\theta_{x'}) \, \diff x' \nonumber\\
\qquad {} + \frac{8}{1-x} \int^{1}_{x} \frac{1}{(1+x')^2} d^\ell_{2,2}(\theta_{x'}) \, \diff x'\, . \label{eq:dminus2fromd2}
\end{gather}
From the symmetry $d^{\ell}_{2,-2}(\pi - \theta) = (-1)^{\ell} d^{\ell}_{2,2}(\theta)$ we then immediately obtain the inverse relation
\begin{gather}
d_{2,2}^\ell(\theta_{x}) =  d_{2,-2}^\ell(\theta_{x}) \nonumber\\
\qquad {} - \frac{8(2-x)}{(1+x)^2} \int_{-1}^{x} \frac{1+x'}{(1-x')^2} d_{2,-2}^\ell(\theta_{x'}) \, \diff x' \nonumber\\
\qquad {} + \frac{8}{1+x} \int_{-1}^{x} \frac{1}{(1-x')^2} d^\ell_{2,-2}(\theta_{x'}) \, \diff x'\, . \label{eq:d2fromdminus2}
\end{gather}

\section{Purification} \label{sect: Purification}
\begin{figure}
    \centering
    \includegraphics[width=1\linewidth]{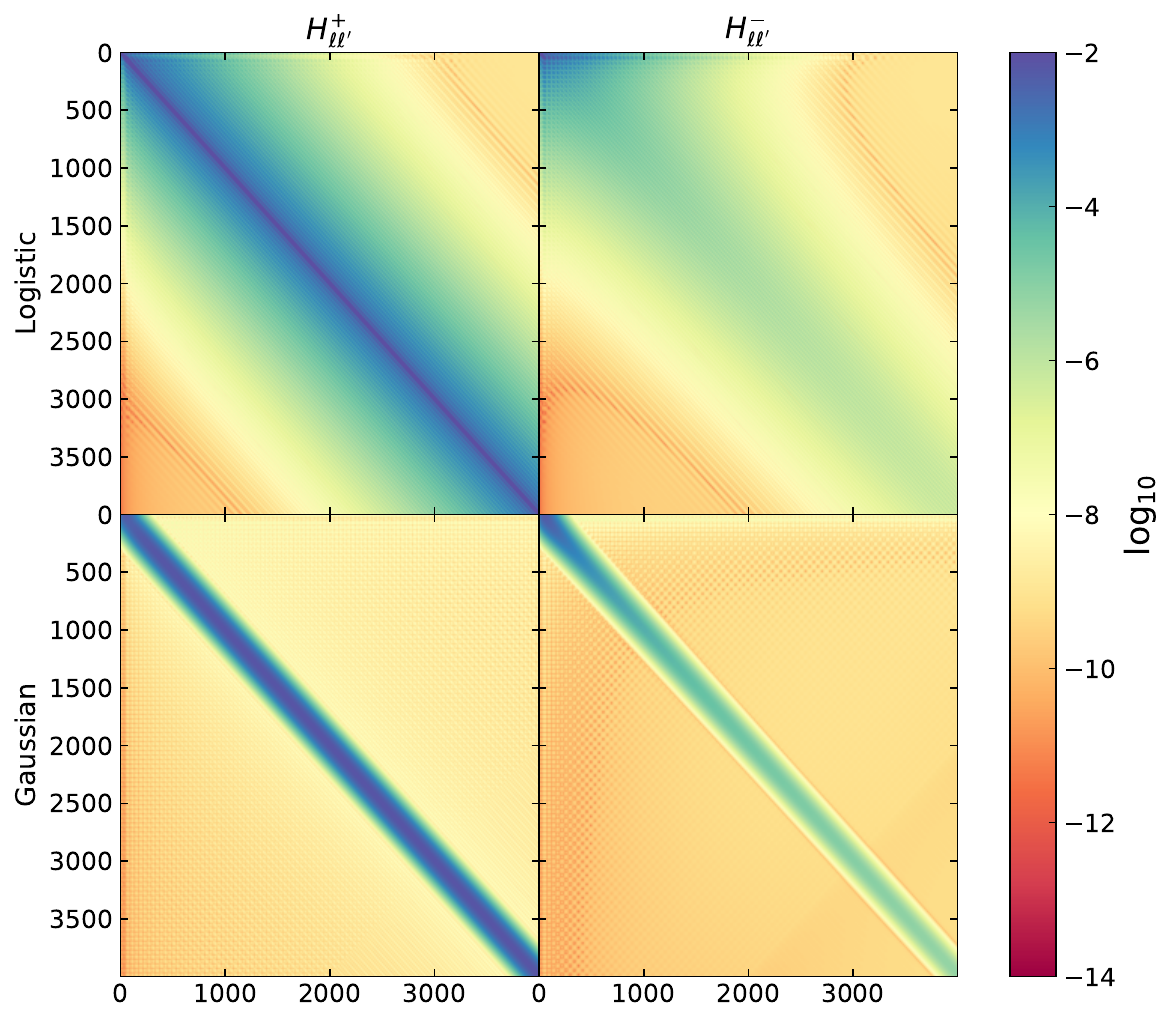} 
    \caption{Apodisation kernels associated with a logistic (top) and Gaussian (bottom) apodisation function, Eqs.~(\ref{eq: logistic_apo}, \ref{eq: gaussian_apo}). Both apodisation functions used a maximum angular separation of $\theta^*=4.9$\,deg with steepness $\lambda_{\rm gaussian} = 24\ln 2$ and $\lambda_{\rm logistic} = 2$, respectively.}%
    \label{fig:H_kernels}%
\end{figure}

\begin{figure}
    \centering
    \includegraphics[width=1\linewidth]{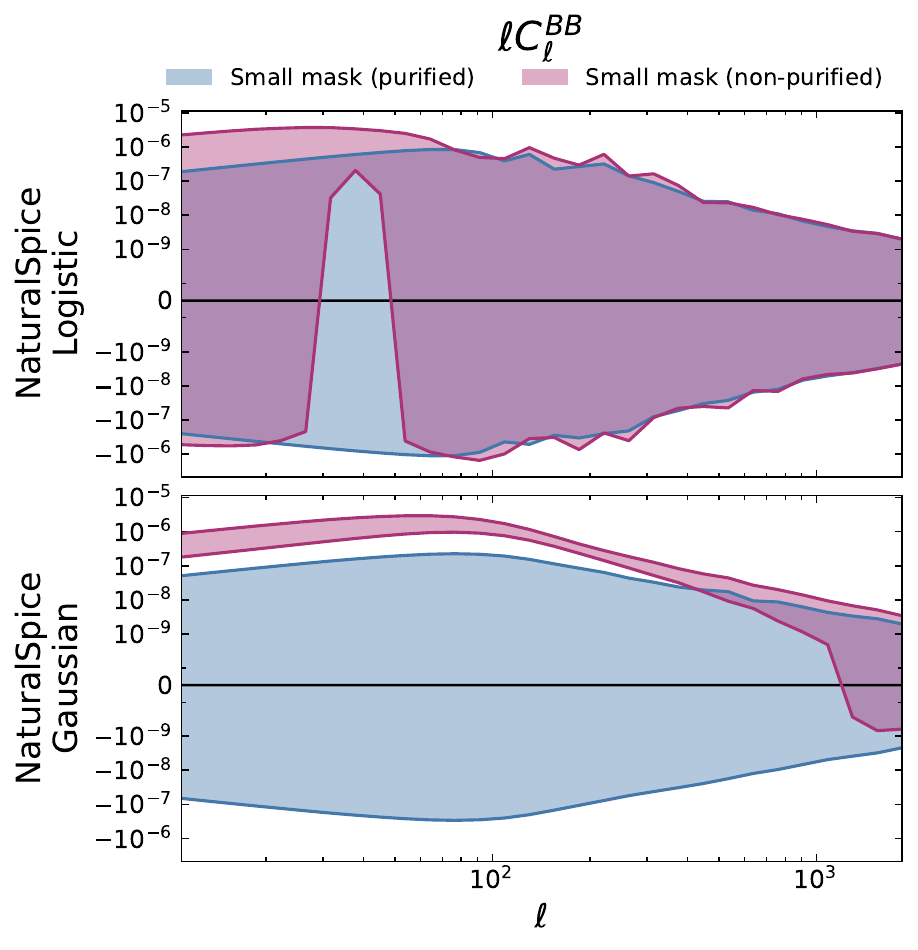} 
    \caption{Impact of purification on the $B$-mode auto-correlation spectrum of the small ($\sim 20 \sqdeg$) mask when using a logistic (top) and Gaussian (bottom) apodisation function. Blue contours show the same purified constraints as in Fig.~\ref{fig:cls} while purple contours show the equivalent constraints but using a naive, not-purified estimator. Purification is essential to remove biases due to $E$-mode power leakage, especially when using broad apodisation functions.}%
    \label{fig:purification}%
\end{figure}

As discussed in Sect.~\ref{Sect: PolSpice} apodisation breaks the orthogonality of the Wigner $d$-matrices, mixing power between $E$- and $B$-mode spectra when estimated naively following Sect.~\ref{sect: NaturalSpice}. Let us exemplify this by explicitly propagating the impact of apodisation in  Eq.~\eqref{eq: pseudo-All} to the final angular power spectra estimates. For the sake of argument, let us focus on the $B$-mode auto-spectrum,
\begin{gather}
\estimator{C}^{BB'}_{\ell} = \pi \bigg[ \int^{\pi}_{0} h(\theta) \, \estimator{\xi}^{+} \, d^\ell_{22}(\theta) \, \sin\theta \, \diff\theta \nonumber \\
- \int^{\pi}_{0} h(\theta) \, \estimator{\xi}^{-} \, d^\ell_{2-2}(\theta) \, \sin\theta \, \diff\theta \bigg] \; .
\end{gather}
Now expand the correlation functions in terms of the expectation values of their associated angular power spectra,
\begin{gather}
\estimator{C}^{BB'}_{\ell} = \sum_{\ell'} \frac{2\ell'+1}{2} \bigg[\frac{H_{\ell\ell'}^{+} + H_{\ell\ell'}^{-}}{2}  \langle C_{\ell'}^{BB} \rangle +  \frac{H_{\ell\ell'}^{+}  - H_{\ell\ell'}^{-}}{2} \langle C_{\ell'}^{EE} \rangle \bigg] \; ,
\end{gather}
where we have defined the kernels
\begin{equation}
    H_{\ell\ell'}^{\pm} = \frac{2\ell'+1}{2} \int_0^\pi h(\theta) \, d^\ell_{2\pm2}(\theta) \, d^{\ell'}_{2\pm2}(\theta) \, \sin\theta \, \diff\theta \; .
\end{equation}
Hence, we can see that apodisation mixes power between the components of the shear field. This is particularly relevant when estimating the $B$-mode auto-spectrum since the $E$-mode auto-spectrum tends to be several orders of magnitude larger than the $B$-mode auto-spectrum, leading to meaningful biases. As discussed in Sect. \ref{Sect: PolSpice}, by constructing pure $E$- and $B$-mode correlation functions, the purified estimators, Eqs.~(\ref{eq:Cl_EE_pure}) and (\ref{eq:Cl_BB_pure}), prevent apodisation from mixing power across modes. Expanding these pure correlation functions in terms of the expectation values of their associated angular power spectra we obtain
\begin{equation} \label{eq:Cl_EE_pure_bias}
    \estimator{C}_\ell^{EE'}  = \sum_{\ell'} H^-_{\ell \ell'} \langle C_{\ell'}^{EE} \rangle \; ,
\end{equation}
and
\begin{equation} \label{eq:Cl_BB_pure_bias}
    \estimator{C}_\ell^{BB'}  = \sum_{\ell'} H^-_{\ell \ell'} \langle C_{\ell'}^{BB} \rangle \; ,
\end{equation}
which, while still biased, do not mix power across modes. In Fig.~\ref{fig:H_kernels}, we show the form of these kernels for the logistic and Gaussian apodisation functions considered in this work. We use a maximum angular separation of $\theta^*=4.9$\,deg for both apodisation functions with steepness given by $\lambda_{\rm gaussian} = 24\ln 2$ and $\lambda_{\rm logistic} = 2$, respectively. In this figure we can see the logistic apodisation function leads to broader kernels than its Gaussian counterpart. We can also see that $H^-_{\ell\ell'}$ induces less mixing of power than $H^+_{\ell\ell'}$. This means that the purified $\estimator{C}_\ell^{BB'}$ estimator is meaningfully less biased, even when mixing is not taken into account, than the naive estimator.

For the masks considered in this work, we find that only the smallest $\sim 20 \sqdeg$ mask suffers from statistically significant apodisation biases in its pseudo-$C_\ell$. We show this effect in Fig.~\ref{fig:purification}. In this figure we show that the non-purified $C^{BB'}_\ell$ estimates show significant biases both when a logistic and a Gaussian apodisation function is used. This stresses the importance of purification to obtain unbiased estimates, particularly at low-$\ell$.

\section{Shot noise} \label{sect: shot noise}
So far in this work we have measured angular power spectra from idealised synthetic maps where the positions and shapes of galaxies at any given pixel are perfectly known. However, in real galaxy surveys only a finite number of galaxies are observed. Moreover, in the case of fields of spin-weight 2, only a fraction of the observed ellipticities correspond to cosmic shear due to the presence of intrinsic ellipticities. Thus, the observed ellipticities can be written as
\begin{equation}
    (e_1, e_2) = (\gamma_1 + n_1, \gamma_2 + n_2) \, ,
\end{equation} 
where $(\gamma_1, \gamma_2)$ is the underlying shear field, and $(n_1, n_2)$ is the associated shape noise field. We define the shape noise field such that $\langle n_1 \rangle = \langle n_2 \rangle = 0$ and $\langle n_1^2 \rangle = \langle n_2^2 \rangle = \sigma_e^2$. These noise contributions manifest as an additive bias, $b$, in the measured angular power spectra that can be computed from the mean variance of the measured ellipticities as
\begin{equation} \label{eq: b}
    b = \frac{f_{\rm sky}}{2 \bar{n} } \left(\langle e_1^2 \rangle + \langle e_2^2 \rangle \right) \, ,
\end{equation}
where $f_{\rm sky}$ is the sky fraction observed, and $\bar{n}$ is the number density of galaxies per steradian. Since the two contributions are uncorrelated, the total bias of Eq.~\eqref{eq: b} can be split into a contribution due to the intrinsic noise in the shear field and a shape noise  contribution, $b = b_\gamma + b_n$, where 
\begin{equation}
b_\gamma = \frac{f_{\rm sky}}{2 \bar{n} } \left(\langle \gamma_1^2 \rangle + \langle \gamma_2^2 \rangle \right)
\end{equation}
and
\begin{equation}
b_n = \frac{f_{\rm sky}}{\bar{n}} \sigma_{e}^2    \, .
\end{equation}
 
\texttt{Heracles} removes the total bias by default from the measured angular power spectra. This introduces challenges when comparing them to theoretical expectations because the theory angular power spectra receive a bias contribution from the intrinsic noise of the field but not from the additional shape noise present in observations. \citet{Heracles} show that this difference can effectively be accounted for by considering the so-called `reduced' mixing matrix that satisfies
\begin{equation}
\label{eq:ev-cl-ff-noi}
    \langle \tilde{C}_\ell^{ff'} - b \rangle
    = \sum_{\ell_1} \mathscr{M}_{\ell\ell_1}^{ww'} \, \langle C_{\ell_1}^{ff'} \rangle
     \;.
\end{equation}
 
Thus, when computing estimates of the full-sky angular power spectra from catalogue measurements the \texttt{NaturalSpice} algorithm has to be adapted to invert the impact of the reduced mixing matrix, as opposed to the full mixing matrix. When working with finite catalogues, the angular power spectra of the mask also receive an extra shot noise contribution. Conceptually, this is due to the mask no longer being a binary map but, instead, being a series of delta functions at the positions of the galaxies in the catalogue. Computing the mixing matrix from the mask power spectra after subtracting the shot noise bias yields the reduced mixing matrix. Similarly, the \texttt{NaturalSpice} methodology can be adapted to catalogue analysis by computing the correlation function of the mask from the same bias-subtracted mask angular power spectra. 
 
\begin{figure}
    \centering
    \includegraphics[width=1\linewidth]{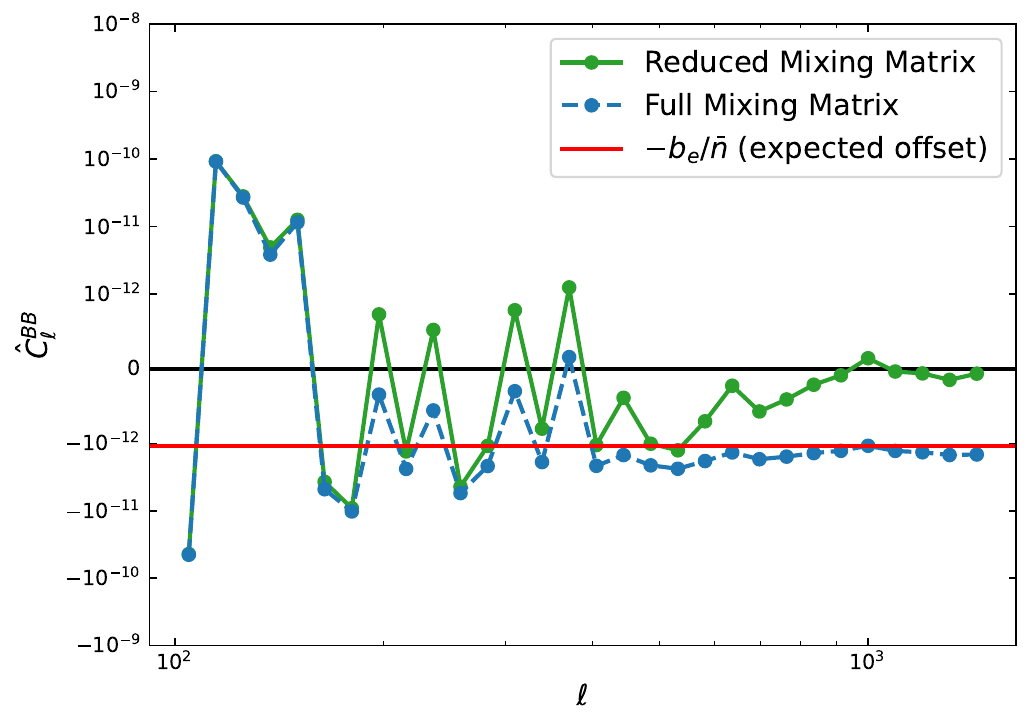} 
    \caption{ \texttt{NaturalSpice} estimate of $C_\ell^{B\!B}$ in the analysis of an idealised catalogue with a finite number of sources but no shape noise in the measured ellipticities. This exercise demonstrates that angular power spectra estimated from finite observations overestimate the amount of bias that needs to be subtracted. This small error is overpowered by the shape noise contribution in realistic catalogues, but it can be important when computing estimates of quantities, such as $\estimator{C}_\ell^{B\!B}$, whose expectation value is zero. However, this bias can be completely accounted for by subtracting the shot noise contribution in the angular power spectra of the mask (associated with inverting the effect of the reduced mixing matrix, green) as opposed to naively inverting the full mixing matrix (blue). }%
    \label{fig:shot_noise}%
\end{figure}
 
In realistic galaxy surveys, the bias contribution from shape noise overpowers the intrinsic bias by orders of magnitude. Thus, the effect of not accounting for the latter can only be appreciated when thousands of simulated fields are considered \citep{Heracles}. However, this small offset can manifest itself when computing the full-sky estimate of $\estimator{C}^{BB}_\ell$ due to its expectation value being centred around zero. This is exemplified in Fig.~\ref{fig:shot_noise} where, to illustrate our point, we consider an idealised catalogue with a finite number of galaxies but no shape noise. In this figure we show how failing to remove the shot noise contribution to the mask angular power spectra (which corresponds to naively inverting the full mixing matrix) causes the \texttt{NaturalSpice} estimate of $C_\ell^{BB}$ to fall beneath zero, as expected from subtracting too much noise. Instead, correctly subtracting this bias (i.e. inverting the reduced mixing matrix) yields the expectation from the theory. 
 
Moreover, it can be shown that this bias is exactly what we would expect from the intrinsic level of noise in the shear field, $b_\gamma$. As previously described, it is not possible to estimate $b_\gamma$ from the data due to the conflation of different sources of noise. Instead, $b_{\rm \gamma}$ can only be estimated from the theory expectations themselves. This can be done by leveraging the fact that for an isotropic and homogeneous field the variance of the field is given by
\begin{equation}
    \frac{1}{2}(\langle \gamma_1^2 \rangle + \langle \gamma_2^2 \rangle) = \sum_\ell \frac{2\ell+1}{4\pi} C_\ell^{E\!E} \, ,
\end{equation}
where $C_\ell^{E\!E}$ is the full-sky theoretical expectation for the auto-correlation spectrum of the $E$-mode. 
 
Finally, we note that the \texttt{NaturalSpice} implementation within \texttt{Heracles} automatically subtracts the shot noise contribution from the mask angular power spectra. Thus, the algorithm can be used out of the box to process catalogues. However, users seeking to write their own implementations should be cautious of the considerations above.

\end{appendix}

%\label{LastPage}
\end{document}